\documentclass[sigconf,nonacm,pdfa]{acmart}

\usepackage[normalem]{ulem}
\usepackage[a-2b]{pdfx}
\usepackage{epstopdf}
\usepackage{tikz}
\usepackage{pifont}
\usepackage{epsfig,endnotes}
\usepackage{amsmath}
\usepackage{color}
\usepackage[titletoc,toc,page]{appendix}
\usepackage{textcomp}
\usepackage{multirow}
\usepackage{graphicx}
\usepackage{fontawesome}
\usepackage{xcolor}
\usepackage{enumitem}
\usepackage{booktabs}
\usepackage{algorithmicx}
\usepackage[ruled,linesnumbered]{algorithm2e}
\usepackage{algpseudocode}
\usepackage{subfigure}
\usepackage{enumerate}
\usepackage{makecell}
\usepackage{balance}
\usepackage{filecontents}
\usepackage{xspace}
\usepackage{caption}
\usepackage{fancyhdr}
\usepackage{setspace}
\usepackage{hyperref}
\usepackage{libertine}
\usepackage{listings}
\usepackage{environ}

\newcommand{\sys}{Outback\xspace}

\newcommand{\red}{}
\newcommand{\blue}{\textcolor[rgb]{0,0,1}}

\newcommand{\todo}[1]{{\it \color{red}\{TODO: #1\}}}
\newcommand{\minghao}[1]
{\noindent{}}

\newif\ifshowheiner
\newcommand{\heiner}[1]{\ifshowheiner\noindent{\textcolor{teal}{\bf \fbox{HL} {\it#1}}}\fi}

\newif\ifshowliuyi 
\newcommand{\liuyi}[1]{\ifshowliuyi\noindent{\textcolor{red}{\bf \fbox{YL} {\it#1}}}\fi}

\newcommand\vldbdoi{10.14778/3705829.3705849}
\newcommand\vldbpages{335-348}
\newcommand\vldbvolume{18}
\newcommand\vldbissue{2}
\newcommand\vldbyear{2024}
\newcommand\vldbauthors{\authors}
\newcommand\vldbtitle{\shorttitle} 
\newcommand\vldbavailabilityurl{https://github.com/yliu634/outback}
\newcommand\vldbpagestyle{empty}

\begin{document}

\title{Outback: Fast and Communication-efficient Index for Key-Value Store on Disaggregated Memory}

\author{Yi Liu}
\affiliation{%
  \institution{University of California Santa Cruz}
  \streetaddress{1156 High Street}
 \country{}
  \postcode{95064}
}
\email{yliu634@ucsc.edu}
\author{Minghao Xie}
\affiliation{%
  \institution{University of California Santa Cruz}
  \streetaddress{1156 High Street}
 \country{}
  \postcode{95064}
}
\email{mhxie@ucsc.edu}
\author{Shouqian Shi}
\affiliation{%
  \institution{University of California Santa Cruz}
  \streetaddress{1156 High Street}
 \country{}
  \postcode{95064}
}
\email{sshi27@ucsc.edu}
\author{Yuanchao Xu}
\affiliation{%
  \institution{University of California Santa Cruz}
  \streetaddress{1156 High Street}
 \country{}
  \postcode{95064}
}
\email{yxu314@ucsc.edu}
\author{Heiner Litz}
\affiliation{%
  \institution{University of California Santa Cruz}
  \streetaddress{1156 High Street}
  \country{}
  \postcode{95064}
}
\email{hlitz@ucsc.edu}
\author{Chen Qian}
\affiliation{%
  \institution{University of California Santa Cruz}
  \streetaddress{1156 High Street}
  \country{}
  \postcode{95064}
}
\email{qian@ucsc.edu}

\begin{abstract}
\heiner{Title is a bit long. It should be disaggregated KVS (not the) as there exist more than one KVS implementations. If you say decoupling, you must say decoupling of A and B. There have to be at least 2 things that are decoupled.}
\liuyi{Fixed it, we will formulate a new one after revising paper.}

Disaggregated memory systems achieve resource utilization efficiency and system scalability by distributing computation and memory resources into distinct pools of nodes. RDMA is an attractive solution to support high-throughput communication between different disaggregated resource pools. However, existing RDMA solutions face a dilemma: one-sided RDMA completely bypasses computation at memory nodes, but its communication takes multiple round trips; two-sided RDMA achieves one-round-trip communication but requires non-trivial computation for index lookups at memory nodes, which violates the principle of disaggregated memory. This work presents \sys, a novel indexing solution for key-value stores with a one-round-trip RDMA-based network that does not incur computation-heavy tasks at memory nodes. \sys is the first to utilize dynamic minimal perfect hashing and separates its index into two components: one memory-efficient and compute-heavy component at compute nodes and the other memory-heavy and compute-efficient component at memory nodes. 
We implement a prototype of \sys and evaluate its performance in a public cloud. The experimental results show that \sys achieves higher throughput than both the state-of-the-art one-sided RDMA and two-sided RDMA-based in-memory KVS by 1.06-5.03$\times$, due to the unique strength of applying a separated perfect hashing index. 

\end{abstract}


\maketitle

\pagestyle{\vldbpagestyle}
\begingroup\small\noindent\raggedright\textbf{PVLDB Reference Format:}\\
\vldbauthors. \vldbtitle. PVLDB, \vldbvolume(\vldbissue): \vldbpages, \vldbyear.\\
\href{https://doi.org/\vldbdoi}{doi:\vldbdoi}
\endgroup
\begingroup
\renewcommand\thefootnote{}\footnote{\noindent
This work is licensed under the Creative Commons BY-NC-ND 4.0 International License. Visit \url{https://creativecommons.org/licenses/by-nc-nd/4.0/} to view a copy of this license. For any use beyond those covered by this license, obtain permission by emailing \href{mailto:info@vldb.org}{info@vldb.org}. Copyright is held by the owner/author(s). Publication rights licensed to the VLDB Endowment. \\
\raggedright Proceedings of the VLDB Endowment, Vol. \vldbvolume, No. \vldbissue\ %
ISSN 2150-8097. \\
\href{https://doi.org/\vldbdoi}{doi:\vldbdoi} \\
}\addtocounter{footnote}{-1}\endgroup

\ifdefempty{\vldbavailabilityurl}{}{
\vspace{.3cm}
\begingroup\small\noindent\raggedright\textbf{PVLDB Artifact Availability:}\\
The source code, data, and/or other artifacts have been made available at \url{\vldbavailabilityurl}.
\endgroup
}

\vspace{-.5ex}
\section{Introduction}
\label{sec:intro}
\vspace{-0.5ex}

Disaggregated memory systems~\cite{case,tutorial,fusee,shan2022towards,flexchain,rethinking,dex,pang} represent a transformative departure from traditional computing architectures, distributing memory storage and computational resources into distinct pools of nodes -- compute pools include nodes that carry rich CPU resources, and memory pools include nodes that carry rich DRAM and storage resources. This framework is prevalent in contemporary data centers and cloud infrastructures~\cite{redy,zhang2020understanding}, providing benefits such as enhanced resource utilization efficiency and flexibility to scale the system out by deploying more hardware. Disaggregated memory systems can harness Remote Direct Memory Access (RDMA)-capable networks~\cite{farm,clover,dinomo,fusee,redn}, featuring substantial throughput capacities (ranging from 40 to 400 Gbps) and small latency within the microsecond range. 
Memory-intensive applications, such as transaction systems~\cite{ford,drtmr,drtmh,motor} and key-value stores (KVSs)~\cite{race,rolex,herd,farm}, 
store the data and index data structures at the memory nodes and perform computation tasks at the compute nodes. 

\heiner{I am not used to these definitions. Often you have a say MySQL database and a memory-based Memcached KV-cache in front. Do you refer to such a setup or are you assuming that there is a 2nd level of caching. E.g. the compute nodes cache some data from the Memcached cache?}\liuyi{We will explain the definitions of disaggregated pools in the first sentence of introduction after bring it out}

\begin{figure}[!t]
\centering
\subfigure[An example of one-sided RDMA.]{
    \label{fig:intro:a}
    \includegraphics[width=0.22\textwidth]{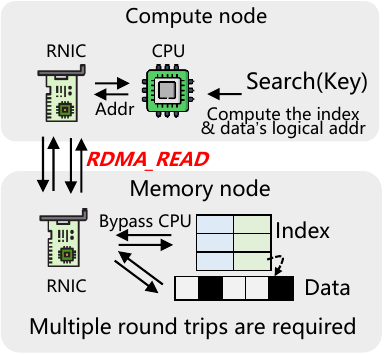}}
\hspace{-1ex}
\subfigure[An example of two-sided RDMA.]{
    \label{fig:intro:b}
    \includegraphics[width=0.23\textwidth]{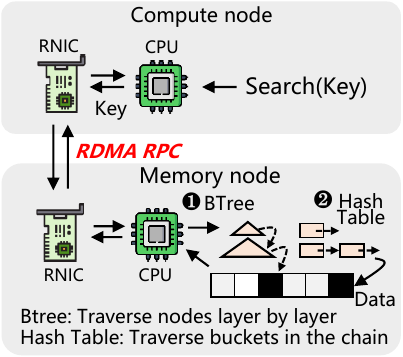}}
\vspace{-3.5ex}
\caption{Examples of two types of RDMA systems.}
\vspace{-4.5ex}
\label{fig:intro}
\end{figure}

Existing RDMA networks for disaggregated memory can be categorized into two types. 1) One-sided RDMA~\cite{sherman,race,cell,rolex,smart} as shown in Fig.~\ref{fig:intro:a}. This type of network completely 
separates computation and memory access tasks. Each data request requires multiple round trips of communication between the compute node and the memory node. At least two round trips are necessary: one to access the index and the other to access the stored data. Note that many indices require multiple layers of accesses~\cite{cell,sherman}, hence they need much more than two round trips~\cite{pilaf,cell}. 
2) Two-sided RDMA or RDMA RPC~\cite{fasst,guidelines}, as depicted in Fig.~\ref{fig:intro:b}, involves computation tasks on both compute and memory nodes, requiring only a single round-trip communication for each request. However, two-sided RDMA cannot bypass the CPU on the memory node, necessitating the CPU on the memory node to execute the computation of the index structure, such as hash computations and key comparisons. Since the CPU resource on a memory node is very limited in disaggregated systems, this design may lead to CPU bottlenecks and potentially higher latency compared to one-sided RDMA~\cite{hstore,sherman}.

A natural question arises: "Can we design a one-round-trip RDMA-based network that does not incur computation-heavy tasks on memory nodes?" 
Achieving this goal is extremely challenging because putting the index on memory nodes leads to CPU bottlenecks while putting the index on compute nodes causes memory bottlenecks and consistency issues. 

This paper presents the first solution to this research problem.
Our key innovation is to design and implement an RDMA RPC-based system, called \sys, which decouples its index into two components. The first component is memory-efficient and includes most computation operations of the index, which is placed onto the compute nodes. The second component contributes to the most memory cost of the index, but its computation is trivial, and it is on the memory nodes. 
Such a design principle of decoupling the index is ideal for disaggregated memory systems: all computation tasks for \texttt{Get} requests and the majority of computation for data \texttt{Insert} requests are offloaded on compute nodes, while memory nodes focus on providing service for memory read and write. Hence, this approach is particularly effective for real-world workloads dominated by \texttt{Get} requests. It is also well-suited for emerging disaggregated memory systems equipped with SmartNICs with limited computation resources~\cite{bluefield, alveo, stingray}.

Similar to prior one-round-trip RDMA networks~\cite{fasst,guidelines}, \sys also relies on two-sided RDMA. 
We implement \sys as a distributed KVS application. The index design of \sys is motivated by a recent advance of dynamic minimal perfect hashing (DMPH), called Ludo hashing~\cite{ludo}. 
The original design of Ludo hashing did not decouple the index into computation-heavy and memory-heavy components, but its perfect hashing property offers the opportunity for a novel decoupling approach that allows data \texttt{Get} requests in one round trip with trivial computation on memory nodes. 
For data \texttt{Insert} requests, we design additional operations to update the index on both the compute and memory nodes to ensure data consistency.

Overall, this paper makes the following contributions:
\vspace{-1ex}
\begin{itemize}
    \item We present a novel solution that provides one-round-trip RDMA with RPC that incurs minimal computation tasks on memory nodes. The design principle of decoupling the index works effectively for emerging disaggregated memory systems.

    \item We design the \sys system as a distributed KVS. We design a decoupled index based on a recent data structure of DMPH. We also designed the algorithms and protocols for supporting data operations and system updates.
    
    \item We implement a prototype of \sys and evaluate the performance on YCSB workloads~\cite{ycsb} and four real-world datasets from SOSD~\cite{sosd}. The experimental results show that \sys achieves higher throughput than both the state-of-the-art one-sided RDMA and two-sided RDMA-based in-memory KVS by 1.06-5.03$\times$.
\end{itemize}

\vspace{-2ex}
\section{Background}
\label{sec:background}
\vspace{-.5ex}

\subsection{Disaggregated Memory with RDMA}

Disaggregated memory systems with RDMA can be categorized into two types: one-sided RDMA systems~\cite{sherman,race,cell,rolex,smart}, and two-sided RDMA (RDMA-RPC) systems~\cite{fasst, guidelines}. An example of one-sided RDMA systems~\cite{sherman,race,cell,rolex,smart} is illustrated in Fig.~\ref{fig:intro:a}. These systems support applications such as KVS and transaction systems with various index data structures, including B/B+ trees, hash tables, radix trees, and learned indexes. However, it is widely recognized that multiple round-trip communications are needed for each \texttt{Get} request: at least one for querying the index and one for reading data. The high communication cost results in both long latency and network congestion. 

\heiner{if you have data to support these claims, provide a forward pointer. E.g. in Section we will show that RDMA introduces X overheads}\liuyi{I think, we can cite the data from cite{scalablerpc} to explain the bad scalability of one-sided RDMA}
\heiner{no need to rely on what other may be concerned about. just say: RPC-based systems introduce CPU overheads...}\liuyi{thanks! ``RPC-based approaches suffer from the large CPU consumption in disaggregated systems, ''}

\begin{figure}[t]
\centering
    \includegraphics[width=0.415\textwidth]{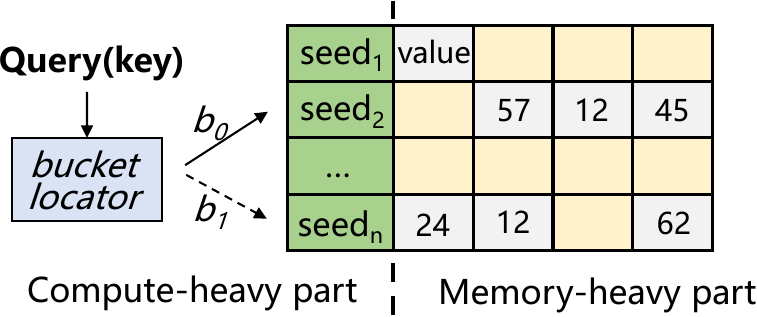}
    \vspace{-1.5ex}
    \caption{Ludo hashing.}
    \label{fig:ludo}
    \vspace{-4.5ex}
\end{figure}

Two-sided RDMA-based systems~\cite{guidelines,fasst} have been investigated to dispatch compute nodes' requests to the memory node via RPC over the RDMA network with only one round trip. As depicted in Fig.~\ref{fig:intro:b}, a data index, such as a B-Tree or hash table, is maintained at the memory node.
When a data query occurs, in addition to polling the RNIC and posting messages, the CPU of the memory node is responsible for traversing the index. The memory node has to perform computational tasks, including hash computation, fingerprint checking, and key comparisons. This process introduces additional computational overhead and memory accesses.
Existing solutions~\cite{race,drtmr,mica} that store keys' fingerprints in their hash tables to save memory usage also introduce extra computation. 
For example, if the memory node employs the state-of-the-art (2,4)-Cuckoo hash table~\cite{cuckoo}, each \texttt{Get} request requires one fingerprint computation and, at most eight rounds of fingerprint checking.

\vspace{-1ex}
\subsection{Dynamic minimal perfect hashing}
\label{subsec:background:dmph}
In this subsection, we first introduce the background of DMPH and then present an existing MPH implementation, Ludo hashing~\cite{ludo}. 

Perfect hashing~\cite{praticalph} represents a family of schemes that designs and manipulates hash algorithms to distribute keys to different buckets in a hash table without collisions.
Since it is impractical to find a single hash function that generates no collisions for a large set of keys, a common approach is to use two levels of mapping. The first level maps keys to a number of groups, each of which contains several keys. The second level addresses key collisions inside each group. Minimal perfect hashing maps $n$ keys to exactly $n$ buckets, but it is inflexible for key insertions and only applicable to a static set. To allow key dynamics, dynamic minimal perfect hashing (DMPH) may use $(1+\epsilon)n$ positions for $n$ keys \cite{scalebricks,ludo}. 
One primary advantage of perfect hashing is that it does not need to store the keys in the hash table. Since perfect hashing eliminates collisions, a key query does not need to compare keys to address collisions. Avoiding storing keys can significantly reduce memory costs, because as a secondary index, the size of keys (usually hundreds of bits) is much longer than the queried value in a hash table (usually a storage address in tens of bits).

One of the most recent solutions of DMPH is called Ludo hashing~\cite{ludo}. 
As shown in Fig.~\ref{fig:ludo}, Ludo hashing~\cite{ludo} first uses a data structure called Othello~\cite{othello}, a dynamic implementation of Bloomier filters \cite{bloomier} with two arrays, as the \textit{bucket locator} to distribute keys into different buckets, each of which includes exactly 4 slots. Then, in each bucket $B_i$, Ludo hashing uses brute force to find a hash seed $s_i$ such that the hash function with $s_i$
can map the 4 keys in the bucket to 4 different slots without collision. Hence, there is no need to store keys in the table for collision resolution. 
The space cost of Ludo is $3.76 + 1.05l$ bits per key, where $l$ is the length of the record value, which is claimed to be the smallest memory cost in the literature~\cite{ludo}. 
\red{The bucket locator leverages Othello arrays~\cite{othello}, which costs 2.33 bits per key. Each bucket contains a 5-bit long seed shared by four keys in Ludo, i.e., 1.25/0.95 bits per key when we set the load factor as 95\%.
Also, the majority of memory cost is for storing the values in the buckets, costing $1.05l$ bits per key.}
We observed that the computation for looking up the slot only needs the bucket locator and the seeds, which are memory efficient. On the other side, the hash table buckets/slots part storing all data values contributed to most memory of this index, but it requires little computation.

\vspace{-2.5ex}
\section{Measurement and Motivation}
\label{sec:motivation}
\vspace{-.5ex}

We wonder if, we remove the computation cost at the memory node, will RDMA-RPC demonstrate much higher throughput than the state-of-the-art one-sided RDMA?
\textbf{If the answer is "Yes", then there is a great opportunity to design a high-throughput RDMA-based KVS by reducing the computation cost at the memory node.} 

Toward this objective, we conduct experiments to analyze the throughput performance of both one-sided RDMA and RDMA-RPC systems with 9 r320 servers in CloudLab~\cite{cloudlab}, each is configured with a Mellanox CX3 adapter (50Gbits).
We compare the performance of the following systems with \texttt{Get}-only workload. (1) RACE hashing~\cite{race}, a state-of-the-art one-sided RDMA-based scheme. Its hashing index is crafted for disaggregated memory, facilitating data retrieval within two round trips. (2) RPC-hash table, a two-sided RDMA method whose compute nodes and memory nodes communicate in RDMA unreliable datagram (UD) mode. Each memory node maintains a chained hash table in its local memory to handle remote data requests. (3) RPC-Dummy. A hypothetical RDMA-RPC method that incurs minimal computation cost at each memory node. 
RPC-Dummy only implements one memory access and then returns any data in the accessed memory at the memory node,  with no extra computation tasks.
RPC-Dummy's throughput can be considered the upper bound among all possible RDMA-RPC systems. We use this method to explore the performance potential of our design objectives.  We vary the number of memory node threads as 1, 2, and 4 in RPC-based approaches, and each memory node thread maintains one Queue Pair (QP) and runs in a distinct CPU core.

\begin{figure}[!t]
\centering
\captionsetup[subfigure]{aboveskip=-2ex}
\vspace{-2ex}
\hspace{-2.5ex}
\subfigure[Throughput of different systems with limited number of memory node threads.]{
    \label{fig:motivation:b}
    \includegraphics[width=0.465\textwidth]{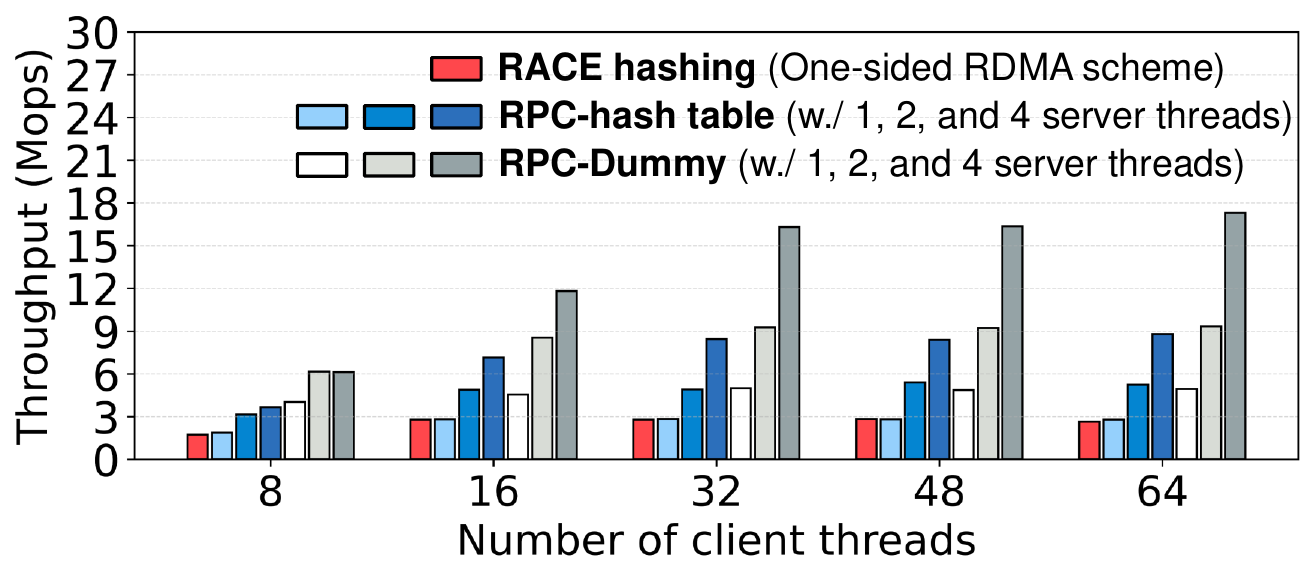}}\\
\vspace{-3ex}
\hspace{-2.2ex}
\subfigure[The CPU time breakdown on a memory node with one thread.]{
    \label{fig:motivation:c}
    \includegraphics[width=0.467\textwidth]{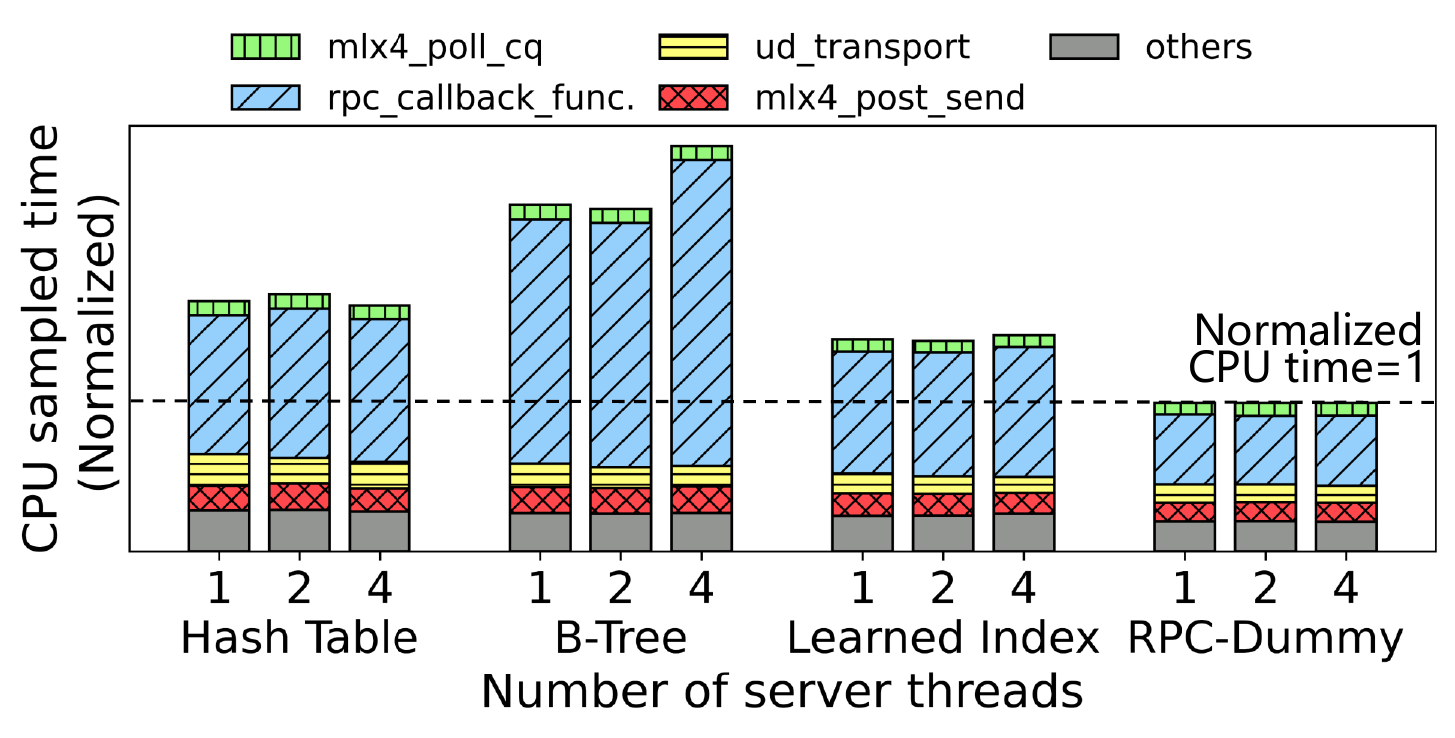}}
\vspace{-3ex}
\caption{Observations from the microbenchmarks.}
\label{fig:motivation}
\vspace{-4.5ex}
\end{figure}

The results are shown in Fig.~\ref{fig:motivation:b}. For one memory node thread (one core), RPC-hash table achieves a throughput similar to that of RACE hashing.
For RACE hashing, multiple reasons limit its throughput, including the two round trips to complete one data \texttt{Get} operation and multiple RC connections of the compute node threads that incur resource contention in the RNIC cache~\cite{scalablerpc}.
RPC-hash table requires only one round trip, but the complexity of querying the index on the memory node introduces extra latency and limits its throughput.  
The throughput of RPC-hash table increases correspondingly when we increase the number of threads to 2 and 4. In contrast, RACE hashing maintains a static performance.
RPC-Dummy can outperform RPC-hash table by around 2$\times$ under the cases of both single and multiple memory node threads. 
Hence an RDMA-RPC network that introduces little computation overhead to the memory node can achieve higher throughput than both existing one-sided RDMA and RDMA-RPC solutions. 
The results suggest that RPC-based KVS has a potential for throughput improvement by reducing computation tasks at memory nodes, which motivates the design of this project.


\textbf{CPU utilization breakdown for RPC-based approaches.}
We run RDMA-RPC with different indices: hash table, Btree, and learned index, at the memory node. 
The CPU time consumed by these four RPC-based KVS systems while handling an equal number of data \texttt{Get} requests is normalized and presented in Fig.~\ref{fig:motivation:c}, with the number of compute node threads fixed at 64. RPC-Dummy takes the least time.
Other approaches consume more time in different amounts.
For RPC-Btree, in addition to the communication overheads for polling $\mathtt{mlx4\_poll\_qp}$ (4.03\%), posting messages $\mathtt{mlx4\_post\_send}$ (7.52\%) and UD transport (6.85\%) from connection management, the most CPU-consuming event is the RPC callback function (70.59\%), which executes local index lookup and data access. 
In all four schemes, the RPC callback function consumes the most CPU time, and the variations in CPU consumption among them are mainly attributed to differences in the RPC callback function. RPC-Btree consumes the most CPU time for RPC callback, followed by RPC-hash table. 
RPC-Dummy spends the least CPU time on the RPC callback function (46.11\%) and serves the most data requests because there is no computation burden for the memory node in RPC-Dummy. 
In disaggregated systems, tasks such as computing hash functions on a hash table, traversing tree nodes in a B-Tree, and executing learned models on a learned index are not ideally suited for memory nodes.
\textbf{The throughput of RDMA-RPC methods is mainly limited by CPU usage during the RPC callback function for index lookups and data reads. High CPU consumption from complex index computations on memory nodes reduces throughput, particularly when CPU resources are constrained, indicating that optimizing these computations can enhance performance.}


\vspace{-1ex}
\section{Design of \sys}
\label{sec:design}
\vspace{-.5ex}

\subsection{Overview}
\label{sec:design:overview}
\heiner{provide brief summary and tie. e.g. In the previous section, we showed that existing KVS suffer from X. We now present..}

Based on the motivation presented in the previous section, we design and implement an RDMA-RPC network that aims to minimize computation tasks on memory nodes, consequently enhancing the system throughput. 
This section presents the design of \sys, a scalable RDMA RPC-based disaggregated KVS that tackles the performance limitations of existing RDMA RPC and one-sided RDMA-based schemes. 
To accomplish this design objective, we decouple the index of \sys into two components: 1) a computation-heavy component running on compute nodes, and 2) a memory-heavy component running on memory nodes.
In particular, DMPH provides an opportunity for this decoupling. By carefully examining the DMPH's read and insertion operations, we observe that the final step consistently is directly retrieving the value from a specific memory location, while all the previous steps are employed to determine that location.
Contrary to DMPH, other hash tables necessitate retrieving the key from the hashed location by key probing and comparison, and only when the key matches the search key, the value can be returned. 
The distinctive process of DMPH motivates us to store all values in the memory-heavy components because they can be read without extra computation. And the steps to determine the location of the value can be placed in the compute-heavy component running on the compute nodes.

\sys requires only a single round trip for data requests while supporting a large number of concurrent compute nodes's requests. In contrast to other RDMA RPC-based approaches~\cite{fasst,herd}, \sys substantially reduces CPU resources required on the memory node.
In the following, we elaborate on the components maintained in the compute pool and memory pool of \sys.

\begin{figure}[t]
    \centering
    \begin{minipage}[b]{0.9\linewidth}
        \includegraphics[width=\linewidth]{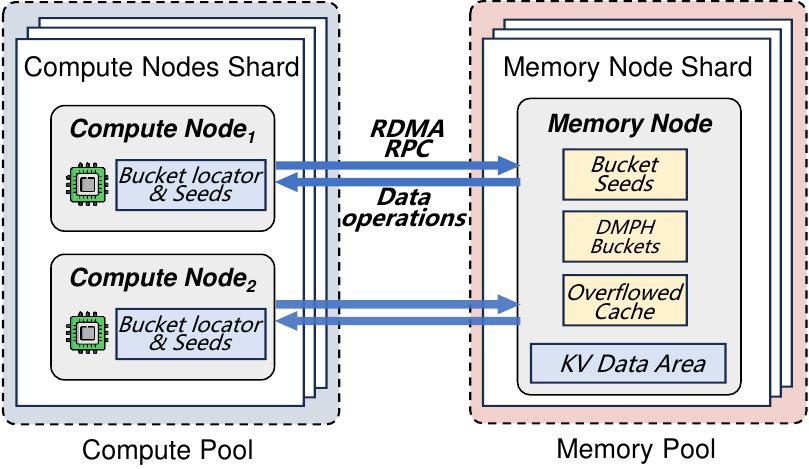}
        \vspace{-3.5ex}
    \end{minipage}
    \vspace{-1ex}
    \caption{\sys overview}
    \vspace{-3.ex}
    \label{fig:overview}
\end{figure}

Fig.~\ref{fig:overview} depicts the overall structure of \sys, which leverages a shared-nothing architecture~\cite{tutorial} for separating data into different shards with consistent hashing~\cite{consistentHashing}. 
The compute pool comprises multiple compute shards, each accommodating several compute nodes.
Note that the configuration for the number of shards and the number of compute nodes depends on the memory budget in compute nodes and the whole size of the datasets. 
For each shard, an index is built based on the keys of the shard, and the returned values of the index represent the memory locations that store the corresponding data associated with the keys. 
The index is decoupled into the compute-heavy and memory-heavy components.  
Each compute node is allocated a memory budget for caching the compute-heavy component, including the bucket locator and the seeds. The default setting is there are 64 million keys in a shard, and the memory overhead on each compute node is less than 50MB (\S\ref{sec:eval:mem}). This is considered a small overhead because recent one-sided RDMA solutions cost over 300 MB on each compute node for index caching and other purposes~\cite{rolex,xtore}. 
All compute nodes in the same shard will connect to the memory node with RDMA RPC for data operations and one-sided RDMA for new bucket locator fetching after index resizing -- the details will be explained in \S\ref{sec:design:resizing}.
Each shard consists of one memory node, which contains the most updated bucket seeds, overflowed cache, DMPH buckets, and KV data in the shard. 
The DMPH buckets store the data addresses in the KV data memory space of the keys in the shard. The latest bucket seeds are maintained to ensure the consistency of data insertion. Additionally, the overflowed cache for KV pairs is used to temporarily hold the pair of the new key and the address, which cannot be inserted into DMPH buckets without the need for hash table resizing. We leverage a hash table to work as the overflowed cache in \sys.
\red{The KV data in each shard is replicated to two other shards, serving as replicas with checkpoints. These two replica shards can be chosen as the two successive shards in the consistent hashing ring. Each key's primary replica shard is referred to as the \textit{primary shard} of the key. Each shard is identified by a \texttt{uuid}. We assume there is a \textit{service layer} in front of the compute nodes responsible for only forwarding data requests to one of the compute nodes in the primary shard based on the key's hash value in the consistent hashing ring. 
After the memory node in the primary shard completes a data update operation, it forwards the update to its replica shards. To ensure load balance among compute nodes within a shard, the service layer maintains a counter for each shard and distributes requests to the compute nodes in a round-robin fashion.}

\begin{figure}[t]
    \centering
    \begin{minipage}[b]{0.9\linewidth}
        \includegraphics[width=\linewidth]{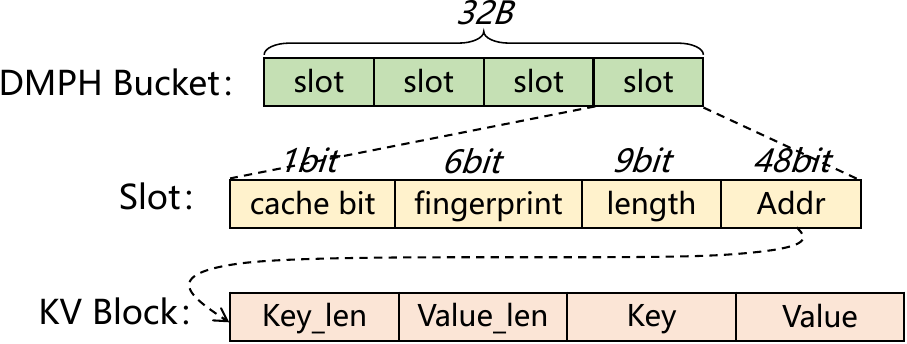}
    \end{minipage}
    \vspace{-1.ex}
    \caption{The data layout in a DMPH bucket.}
    \label{fig:layout}
    \vspace{-3.5ex}
\end{figure}

\begin{figure*}[!t]
\centering
    \subfigure[\texttt{Get} operation.]{
        \label{fig:op:lookup}
        \includegraphics[width=0.335\textwidth]{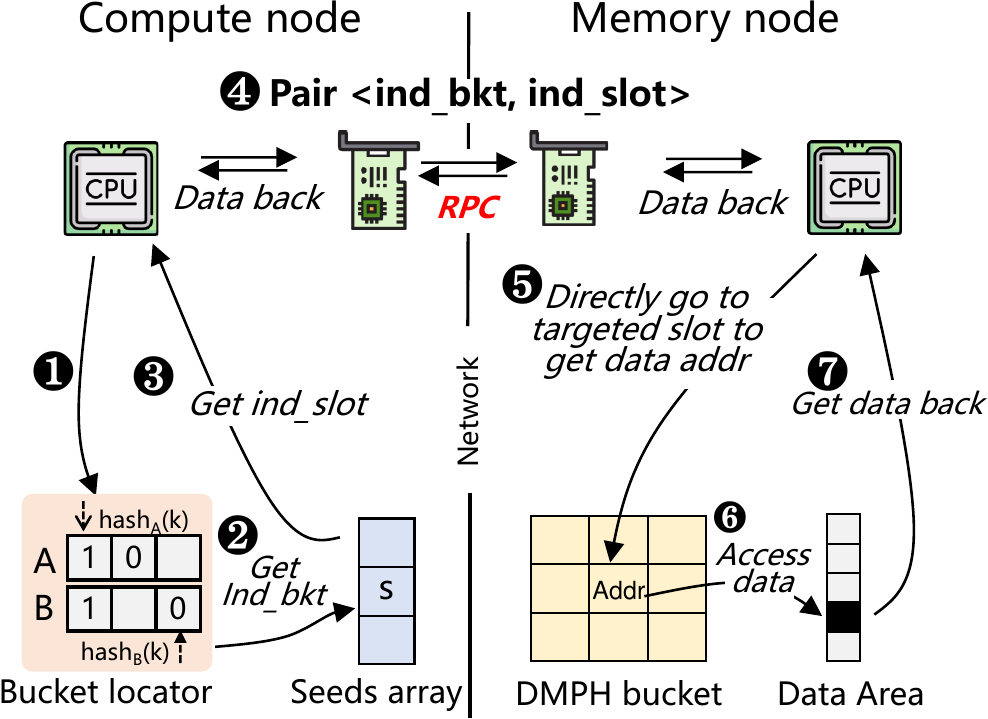}}
        \hspace{-1.2ex}
    \subfigure[\texttt{Insert} operation.]{
        \label{fig:op:insert}
        \includegraphics[width=0.329\textwidth]{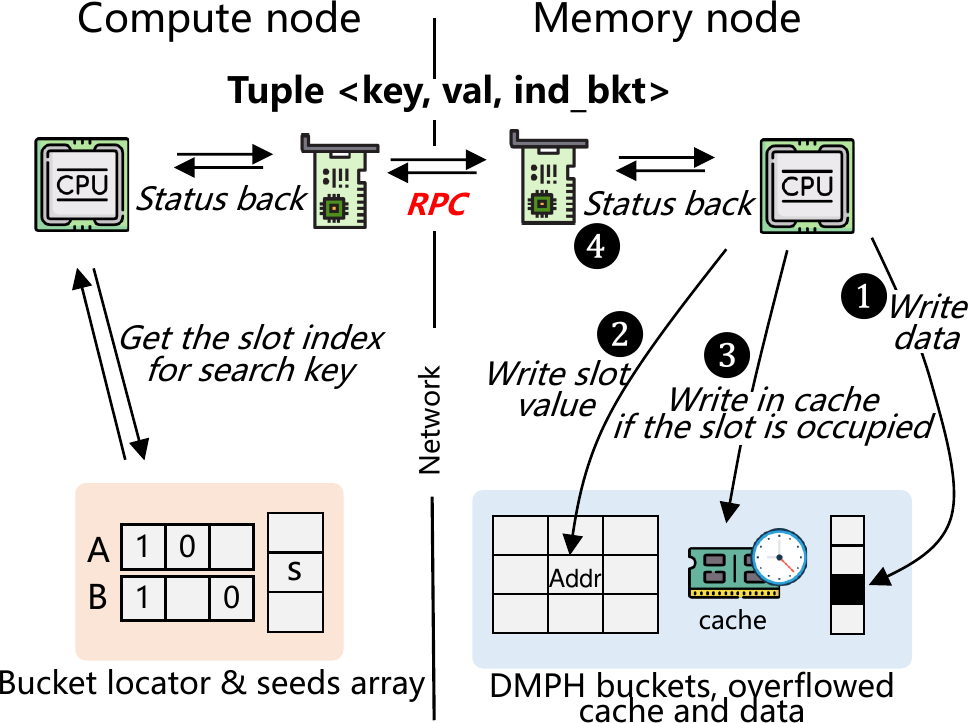}}
        \hspace{-1.2ex}
    \subfigure[\texttt{Update} and \texttt{Delete} operation.]{
        \label{fig:op:update}
        \includegraphics[width=0.32\textwidth]{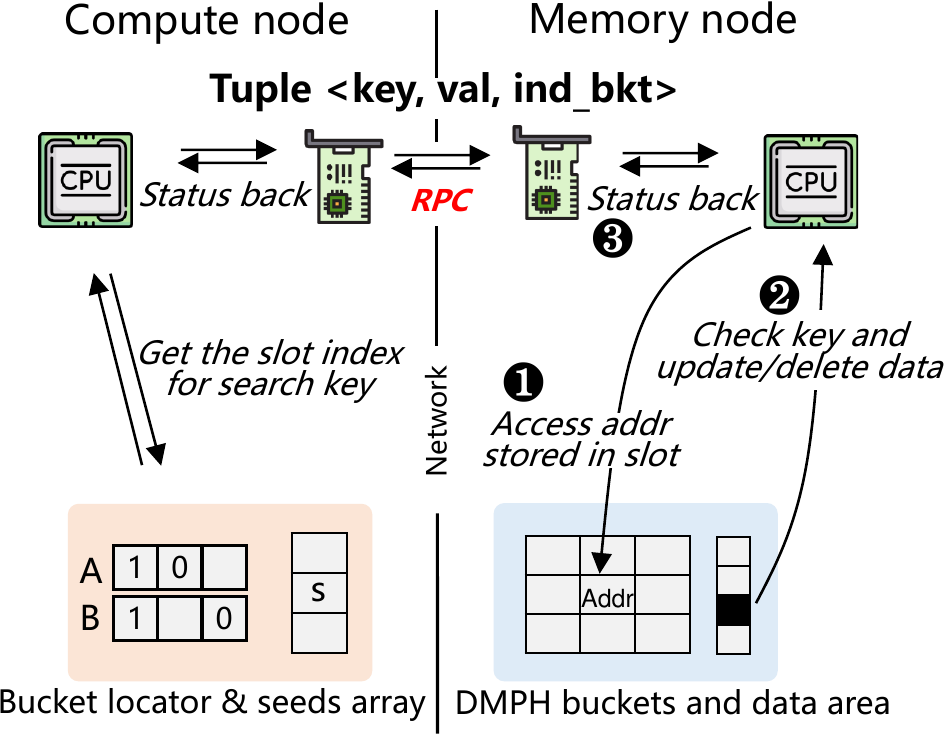}}
\vspace{-3ex}
\caption{Data operation protocols in \sys.}
\vspace{-2.5ex}
\label{fig:op}
\end{figure*}

\vspace{-.5ex}
\subsection{Decoupled DMPH index}
\label{sec:design:table}
In this section, we explain the detailed data structure and its components maintained in the compute node and the memory node. 
We reuse the design of Ludo hashing as introduced in \S\ref{sec:background}. There are two candidate buckets for each key, and the bucket locator runs 
a data structure called Othello~\cite{othello} to determine which bucket the value of the search key is stored in. 
Each Ludo bucket contains one seed and four slots. By computing a hash value with the search key and the seed, the key is mapped to an exact slot of the bucket without colliding with other keys within the same bucket. The value stored in the slot represents the key's data address and is utilized to retrieve the corresponding data.

We decouple the entire data structure of Ludo hashing into two components. The compute-heavy component running on each compute node stores both the bucket locator and the seeds for all DMPH buckets. 
This component completes all computations related to finding the location that stores the value of the search key and costs only $3.76n$ bits -- $2.33n$ bits for the bucket locator and $1.43n$ bits for the seeds, where $n$ refers to the number of KV pairs in a shard.
Within the memory node, the memory-heavy component consists of all DMPH buckets that store the data addresses for all keys in the shard. Assuming the load factor of the DMPH table is set to $\epsilon$ with a default value of 0.95, the number of DMPH buckets will be $n/(4\cdot\epsilon)$ as each bucket accommodates four slots. 
The detailed layout for each DMPH bucket is illustrated in Fig.~\ref{fig:layout}, and each bucket is 32-Byte long with four packed slots. 
There are four fields in each slot: cache bit (1 bit), fingerprint (6 bits), length (9 bits), and data address (48 bits). 
The cache bit serves as an indicator to identify whether another key(s) share the same slot, with its index stored in the overflowed cache. 
Meanwhile, the 6-bit fingerprint is only utilized during the index update process to verify if the KV data referenced by the address in this slot corresponds to the search key or not. This fingerprint check is exclusively applied during data write requests, and any false positives do not impact the final result. This is because a comprehensive recheck of the full key occurs after accessing the actual KV data block on the compute node side.
Note that read requests do not need to check the fingerprint. 
The address signifies the starting offsets of the KV block, while the length indicates the byte length of the entire KV block in the underlying KV data area.
In the underlying data area, the KV block is compactly stored with four fields. The initial two numbers, each occupying 8 bytes, denote the length of the key and the subsequent value field.

The overflowed cache accommodates the key-address pair that cannot be inserted into the mapped DMPH bucket without modifying the bucket locator or resizing the entire hash table.

For an estimation, if $\epsilon=0.95$, the component at the compute node contributes to only 5.5\% of the total memory size of the index while the component at the memory node accounts for the larger portion of 94.5\%.

\vspace{-1.5ex}
\subsection{\sys operations and protocols}
\label{subsec:design:operation}

This subsection presents the data operations and the corresponding protocol of \sys, including the data \texttt{Get}, \texttt{Insert}, \texttt{Update}, and \texttt{Delete} operations, as shown in Fig.~\ref{fig:op}.
\vspace{-1.ex}
\subsubsection{Data Get operation.}
As shown in Fig.~\ref{fig:op:lookup}, the compute node maintains the bucket locator (two Othello arrays $A$ and $B$) and the seed array $s$. Meanwhile, the memory node maintains the DMPH buckets that store KV addresses and the KV data in a disjoint memory area. 
When there is a data \texttt{Get} request for key $k$, the compute node will \ding{182} compute the bucket index from the bucket locator by looking up two bits on the two arrays, respectively. 
Assuming the bucket index that stores the queried key is $ind\_bucket$, the compute node will then proceed to \ding{183} compute the slot number within the bucket with the hash function and the seed $s[ind\_bucket]$. 
At this point, the compute node \ding{184} gets both the bucket index and slot index in the MPH buckets, and it \ding{185} posts them to memory nodes with RDMA\_SEND in the opaque fields. 
After the memory node gets the message and parses the index numbers of the bucket and slot, $ind\_bucket$ and $ind\_slot$, it will \ding{186} go directly to the MPH buckets to access the exact slot without any extra computation. 
Then, the memory node \ding{187} gets the data offset in the underlying KV data area from the last 48-bit field of the slot. At last, \ding{188} the KV data will be read back and returned to the initiating compute node for full key check. 
\red{For example, when a compute node requests data for key 5, it computes the bucket index 10 and slot index 0 based on the bucket locator and the locally stored seeds. Then, the pair of indices (10,0) is sent to the memory node. The data index stored in the indicated slot of the memory node is read, and the corresponding data block is returned. Lastly, the compute node checks the cache bit and a full key to see if the Makeup\texttt{Get} is needed.}

There could be some KV pairs that are temporarily inserted into the overflowed cache during the updates and reconstruction of the index. 
In this circumstance, the compute node is tasked with checking the cache bit, ensuring that the returned full key aligns with the queried one. If the key does not match the requested one, and the cache bit in the slot is set to 1, the compute node will initiate another \texttt{Get} makeup request with the $ind\_slot$ specified as -1, signaling the memory node that the returned key does not match the requested key.
While it is possible to offload the full key comparison task to the memory node, saving one round trip, this approach introduces computation overheads on the limited remote core resources. To make the common case easy, we opt to assign the full key check task to compute nodes.

\textbf{Makeup \texttt{Get} request.}
When the KV data returned to the compute node does not match the requested key, there are two reasons: (1) The requested key is kept in the overflowed cache. The KV pair is inserted after the DPMH table is constructed, and the hashed slot is occupied by another key. 
(2) The requested key is in another slot of the hashed bucket. This case results from changing the order of keys based on the new seed within the bucket when the inserted key can fit into the current DMPH table (detailed in Section~\ref{sec:design:insert}). 
Due to the above two situations, the compute node will send the makeup \texttt{Get} request with the $ind\_slot$ as -1 to the memory node.
The memory node will search the overflowed cache first; if there is a cached item matching the full key of the requested key, it will read the data and return it to the compute node. 
If not, it will read out all the KV blocks referred by the hashed bucket (at most four) and compare the keys until it finds the requested key. 
Additionally, the new seed will be returned back to the compute node if the key is found in another slot, and the compute node will update the copied seeds array for this bucket locally.

\vspace{-1.5ex}
\subsubsection{Data Insert operation.}
\label{sec:design:insert}
The main idea of implementing the data \texttt{Insert} operation of \sys is to determine if we can insert the key into the index without significant changes to the current bucket locator.
If an \texttt{Insert} operation only requires changing the value in one DMPH bucket, \sys can make this change directly. 
However, if a \texttt{Insert} operation will cause the index to resize, which usually happens after a number of insertions, \sys needs to ensure the correctness of the \texttt{Insert} operation and following lookups during index resizing. 
As shown in Fig.~\ref{fig:op:insert}, like \texttt{Get} operation, the compute node will get $ind\_bucket$ and $ind\_slot$ from the bucket locator and the seeds through multiple hashing computations. Different from \texttt{Get}, the RPC message posted to QP should include the full key. 
Thus, the memory node can parse the $ind\_bucket$, $ind\_slot$, and the key from the message and execute the following steps. 
\ding{182} the memory node will write the data into the underlying data area, then it can get the data length and the address (offset in the data area) for indexing. 
After the memory node composes the value from the corresponding slot with the cache bit (set to zero by default), fingerprint, length as well as address, it \ding{183} will try to insert it in the DMPH table. 

We discuss the rest of \texttt{Insert} in three cases:

$\bullet$ \textbf{\texttt{Insert} without bucket locator and seed change.} 
The memory node checks the slot indicated by $ind\_bucket$ and $ind\_slot$. If the length field is empty (length is 0), signifying there is no key associated with this slot, the memory node inserts the composed slot value (Fig.~\ref{fig:layout}) into this location and returns \texttt{SUCCESS} to the compute node. Conversely, if the length is non-zero, indicating that an existing key is using this slot, the memory node proceeds to check the fingerprint and compares the full key to determine if the original key in this slot matches the inserted key. If they match, the insertion is resolved and treated as an \texttt{Update} operation. The fingerprint can prevent the memory node from reading the full key in the KV data area if they are not the same.

$\bullet$ \textbf{\texttt{Insert} with seed changes but the bucket locator remains the same.}
If the key associated with the targeted slot does not match the newly inserted one, an examination is made to determine if there is another available slot within this bucket. Assuming there are only three keys in this bucket, and the slot indicated by $ind\_slot$ is already occupied by a different key, the memory node endeavors to find a new seed that accommodates all four keys in the bucket without causing collisions, thereby preserving perfect hashing policy in this bucket.
The other three keys are read from the underlying KV data area, and the memory node employs a brute-force approach to identify a new seed for perfect hashing within this bucket. Importantly, the bucket locator does not need to change because all four keys remain in the same bucket. Subsequently, the updated seed for this bucket is returned to the compute node, which then propagates this modification to other compute nodes in the same shard. 

$\bullet$ \textbf{Insert data to overflowed cache.}
When all four slots within the bucket are occupied, and the memory node is unable to find an empty slot for the inserted key, the pair of the key and the KV block address will be \ding{184} placed in the overflowed cache. Also, the cache bit in the conflicted DMPH slot will be set to 1 to indicate at least one key in the overflowed cache sharing the same hash slot. 
Instead, when the number of KV pairs in the overflowed cache reaches a predefined threshold, the memory node initiates the index resizing process to accommodate more KV pairs in a new DMPH table.

The data insertion process on each memory node works as follows. 
At first, the memory node will lock the data operations on the targeted bucket
to prevent the potential data operations on this bucket. 
The inserted key might have been stored in the DMPH table before. Thus, the memory node will check if the insert request can be resolved to a data update operation by comparing the fingerprint and the underlying full key. 
Then, the memory node first writes the KV block to the underlying data area 
and processes the data insert request based on the stored bucket keys 
into the mentioned three cases. 
Finally, the memory node unlocks the bucket after it finishes the data insert operation.  
Note that the data insert request tuple sent by the compute node consists of the KV pair and $ind\_bucket$, not including $ind\_slot$. 
The reason is that the memory node keeps the most update seeds array in the shard and can use the seeds to do the hash computation as the slot locator. 
Also, the bucket locator is not maintained in the memory node, and the data insert operation will not modify it after the DMPH table is constructed every time.
This choice is made because modifying the bucket locator requires changing seeds for keys in at least two buckets, leading to more computational overhead. 

\vspace{-2.ex}
\subsubsection{Data Update and Delete operations.}
For data update and deletion, the compute node also acquires the $ind\_bucket$ and $ind\_slot$ from the bucket locator and the seeds array. Like the \texttt{Insert} operation, the compute node transmits the full key to the memory node.
As illustrated in Fig.~\ref{fig:op:update}, the memory node directly accesses the address of the KV data from the DMPH bucket and verifies whether the requested key matches the underlying data. Once the memory node confirms the key, for \texttt{Delete}, it marks the length of the slot value as zero and returns the corresponding status. In the case of \texttt{Update}, it writes the new data to the underlying data area. 
If the cache bit is set to 1 and the keys differ, the memory node will go to the overflowed cache to get the data address.


\vspace{-.5ex}
\subsubsection{\red{Concurrency control.}}
\label{sec:design:concurrency}
\red{Each bucket in the DMPH table within the memory node has a mutex lock. Prior to executing any \texttt{Insert}, \texttt{Update}, or \texttt{Delete} operation, the relevant bucket is locked, blocking any access to its indices. Subsequently, the operation is executed and the lock is released. During the lock period, all other operations targeting this bucket are buffered and only processed once the lock is released.}

\begin{figure}[!t]
    \centering
    \begin{minipage}[b]{0.95\linewidth}
        \includegraphics[width=\linewidth]{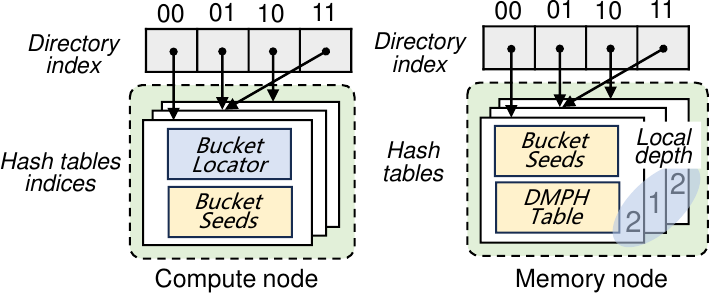}
        \vspace{-5ex}
    \end{minipage}
    \caption{\red{Extendible hashing in \sys.}}
    \vspace{-4ex}
    \label{fig:design:extendible}
\end{figure}

\vspace{-1.ex}
\subsection{Index resizing}
\label{sec:design:resizing}
\red{When the number of KV pairs in the overflowed cache surpasses a predefined threshold, index resizing and reconstruction become necessary to accommodate the KV pairs into a new hash table. This resizing process introduces two challenges: (1) managing data operation requests during resizing and (2) efficiently coordinating the compute node and memory node to transfer the bucket locator and seeds.}

\red{To support data requests on runtime while index resizing, 
\red{we apply extendible hashing~\cite{race,dash} to allocate a new DMPH table to accommodate more keys' indices, and a \textit{directory index} is used to identify the multiple DMPH tables, which is an additional hash layer as shown in Fig.~\ref{fig:design:extendible}.}
This approach reduces the number of keys that need to be moved during index resizing and shortens the resizing duration.
Compute nodes maintain the bucket locator and seeds array for each single hash table, while memory nodes store the most update seeds array and DMPH tables, as well as local depth array~\cite{dash,race}.}


\red{In each shard, we have two size thresholds for overflowed cache; One is for slowing down insertions, $s_{slow}$. The memory node reaching this threshold will enter the index resizing process. The other threshold is the size when the memory node stops any following insertions $s_{stop}$ even if the index resizing is not finished and $s_{stop}>s_{slow}$. 
We set $s_{slow}$ as the load factor of the DMPH table becomes 97\%, or the overflowed cache is filled with half of the size. $s_{slow}$ is set when the overflowed cache is filled with over 90\% space.}

\begin{figure}[!t]
    \centering
    \includegraphics[width=0.93\linewidth]{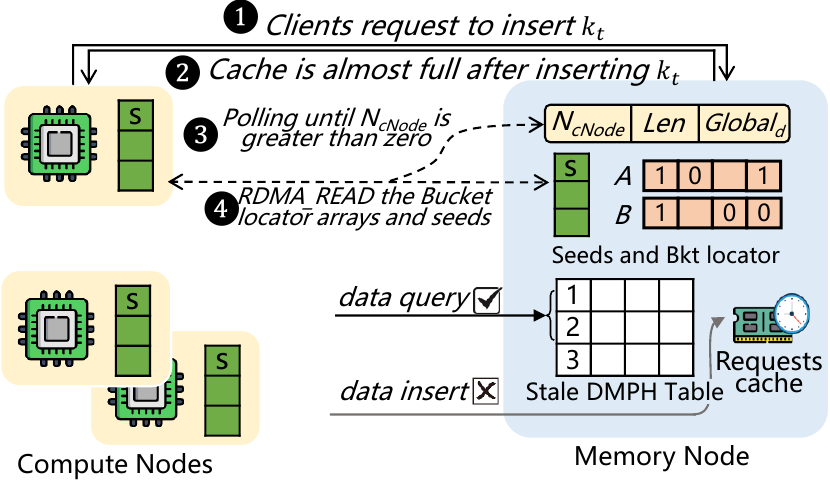}
    \vspace{-2.ex}
    \caption{Index resizing in \sys.}
  \vspace{-4ex}
    \label{fig:resize}
\end{figure}

\red{As shown in Fig.~\ref{fig:resize}, when \ding{182} the overflowed cache size reaches $s_{slow}$ after an \texttt{Insert} request from a compute node, \ding{183} the memory node will return the status \texttt{PRE\_RESIZE} to the compute node, and the compute node will create a new connection manager for preparing and listening to build a one-sided RDMA connection with the memory node. 
The memory node will return \texttt{PRE\_RESIZE} to the data requests for all compute nodes in this shard and count up the number of compute nodes that got the information. After all the compute nodes get it or the overflowed cache size reaches $s_{stop}$, The memory node will build the one-sided RDMA connection (RC) with all compute nodes.
The registered memory area in the memory node consists of five fields: (1) The value of the first eight bytes $N_{cNode}$ indicates the number of compute nodes in this shard, but it is set to zero at the beginning to indicate that the new index has not been completely reconstructed. After it finishes, the value will be set to the number of compute nodes in this shard; (2) the second value of the following eight bytes $len$ refers to the total length of the newly written bucket locator arrays and seeds array; (3) $Global_d$ refers Global depth~\cite{dash} value in current extendible hashing; (4) newly computed seeds array; and (5) bucket locator arrays $A$ and $B$.}

\red{On the compute node, once a connection is established with the memory node, it continuously sends RDMA\_READ requests to retrieve the first two values $N_{cNode}$ and $len$ in the registered memory of the memory node. If $N_{cNode}$ is greater than zero, that means the bucket locator arrays and the seeds array have been successfully constructed and written into the memory area. \ding{184} The compute node then issues another RDMA\_READ requests to fetch all the subsequent $len$ data. Additionally, an atomic primitive of fetch-and-add \texttt{FAA} is executed to decrement $N_{cNode}$ by one, signifying the completion of a compute node fetching the new index data.}

\red{Before the new bucket locator and seeds array is constructed, upon receiving an \texttt{Insert} or \texttt{Delete} request, the memory node returns a \texttt{FALSE} status to compute nodes. Then, the memory node caches the \texttt{Insert}/\texttt{Delete} requests and implements them later after the index data moves to the new DMPH table. 
For \texttt{Get} and \texttt{Update} requests, the memory node will continue serving it on the stale DMPH table. The reason is that no new data insertion would be implemented during resizing, and the keys' $ind\_bucket$ and $ind\_slot$ will not change.}

\red{Once all compute nodes have obtained the new bucket locator arrays and seeds, $N_{cNode}$ in the memory node becomes zero. The memory node detects this change through periodic checks at a frequency of 2 times a second. It proceeds to discard the bucket locator arrays to free up memory space, as they will remain unchanged until the next MPH resizing. 
The memory node will also delete all moved keys in the stale DMPH table by marking the length field as 0.
Then, the reliable connections with all the compute nodes will be terminated by the memory node, and all the compute nodes shift to use both the DMPH tables with the extendible hashing for processing data requests.}

\red{Note that all hash table-based disaggregated KVS require enlargement and shrinking capacity at runtime. The computation time for the extendible hashing layer is the same for \sys and prior works~\cite{race,dash,farm}. In Section~\ref{sec:eval:resizing}, we will show the influence on \sys throughput during index resizing.}

\subsection{Analysis}
\label{subsec:design:analysis}
In this section, we provide the theoretical analysis of the time complexity of the various data operations, as well as the estimation of the memory cost in both compute nodes and memory nodes.

\noindent\textbf{Time complexity.}
For \texttt{Get} operations, each compute node is tasked with determining locations of the DMPH bucket and slot that stores the address of the requested KV. 
This involves two hash computations, namely $hash_A(k)$ and $hash_B(k)$, to access two bits in the bucket locator arrays. Subsequently, an additional hash computation with the bucket seed is performed to locate the specific slot. Then, the memory node can access the slot without further computation and proceed to read data from the referenced KV block.
\red{By default, we use a (2,4)-Cuckoo hash table~\cite{cuckoo} as a fallback table if no seeds can perfectly hash the four elements. In the worst case, accessing the Cuckoo hash table requires two additional hash computations and at most 8 key checks, resulting in a time complexity of O(1) for operations involving the Cuckoo hash table. Therefore, the worst case complexity remains O(1).}
For both the compute node and the memory node, the data \texttt{Get} operation incurs a small constant time. This time complexity extends to data update and data removal operations.

The only difference in \texttt{Insert} lies in the potential time overhead incurred in finding a new seed for the keys in the bucket. To address this, we have set a maximum number of trying times to 256 (8-bit seed). The reason is that we have not encountered a scenario in which no seed can be found within [0, 255] to separate those four keys without collision. We also have a fallback table (storing the key and the KV block address) to deal with rare cases when a group of keys appears that cannot be distributed into distinct slots by MPH. Statistically, we have observed no buckets that cannot be perfectly hashed with a seed length of 8. Therefore, the time cost associated with data insertion is also constant.

\noindent\textbf{Memory usage.}
In compute nodes, the memory usage is allocated to the bucket locator and bucket seeds. According to Ludo~\cite{ludo}, the bucket locator arrays consume 2.33 bits per key. The 8-bit seed is shared among four keys in a bucket. Assuming there are $n$ KV pairs in a shard, with a load factor of $\epsilon$ for the MPH table, the memory cost in a compute node is calculated as $(2.33+2/\epsilon)n$ bits.

In addition to the underlying KV data, memory nodes allocate memory to encompass the latest bucket seeds, DMPH buckets, and the overflowed cache. Each bucket incurs a cost of 32 bytes, and the cache item contains the full key size and the data address. Given a cache size of $m$ and a cache item size of $c$ bits, the overall space budget (in bits) for indexing in a memory node is $66n/\epsilon+m\cdot c$.

\vspace{-1ex}
\subsection{Discussion}
\label{sec:discussion}
\vspace{-.5ex}
\red{\textbf{General applicability on traditional data structures.}
The design principle of \sys can boost data search in traditional data structures with the capability of serving range queries. Specifically, perfect hashing can boost the search process with one-time hash computation with low memory costs that can be cached in compute nodes. For example, the binary search in B/B+ tree leaf nodes can be replaced by perfect hashing computation by searching a seed for hashing keys in leaf nodes.}

\textbf{Ship computation to data.}
\sys decouples the process of DMPH into a memory-heavy component at memory nodes and a compute-heavy component at compute nodes and allows them to communicate via RDMA-RPC primitives. However, the memory accessing based on the given $ind\_bucket$ and $ind\_slot$ still needs a weak power computation unit close to data~\cite{ship}. We can apply \sys to another two promising approaches without using two-sided RDMA verbs. 
\begin{itemize}[left=0em]
    \vspace{-1ex}
    \item \textit{Extended RDMA READ verb.} PRISM~\cite{prism} proposes and simulates an extended one-sided RDMA indirect reading verb \texttt{RDMA\_READ} (\texttt{ptr} \textit{addr}, \texttt{size} \textit{len}, \texttt{bool} \textit{indirect}), where \textit{indirect} indicates if RNIC is supposed to read back the data pointed by the \textit{addr}. This embedded one-sided RDMA verb can free the memory node's CPU and offload the memory reading task in \sys to RNICs. The reason is that \sys can get the exact requested data address without potential data probing.
    \item \red{
    \textit{Performance capacity of Outback with hardware accelerators.} In-network computation~\cite{dinc} has gained attention for accelerating data services in distributed systems by offloading tasks to in-network computation devices~\cite{netsha,cxl-anns} such as SmartNICs/DPUs and CXL~\cite{cxl}. The idea of \sys can reduce the computation burden on SmartNICs by employing one round-trip, one-sided RDMA\_READ. For example, a SmartNIC~\cite{smartnic2,strom,bluefield,ringleader} can be placed on the memory node side, and function as an additional computation unit, and indirect data access tasks can be offloaded to it~\cite{smartnic1}. After the compute nodes in \sys issue a one-sided RDMA to read the queried key's slot and retrieve the address from the DMPH buckets, the SmartNIC can read the memory again via the PCIe switch and obtain the queried data through an additional PCIe round trip. The computation and data search tasks offloaded to the SmartNIC can be alleviated with the assistance of DMPH for the least computation burden.
    }
\end{itemize}

\vspace{-1.5ex}
\noindent\textbf{Shared-nothing architecture.} \sys utilizes a shared-nothing architecture~\cite{tutorial} to prevent the update of cached seeds across compute nodes in different shards. The number of KV pairs in each shard depends on the overall size of the database and the number of shards. A greater number of shards results in fewer KV pairs on each memory node. Consequently, the memory allocation for DMPH seeds and bucket locator on each compute node can be reduced, although additional memory nodes are required. Determining the granularity for sharding KV pairs has always been a tradeoff~\cite{kraska}, and it is recommended to choose the configuration based on the specific application.

\vspace{-2ex}
\section{Performance evaluation}
\label{sec:eval}
\vspace{-1ex}

\begin{figure*}[!t]
\centering
\renewcommand\thesubfigure{}
\subfigure[]{
    \includegraphics[width=0.58\textwidth]{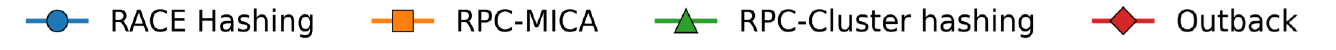}}\\
\vspace{-6ex}
\setcounter{subfigure}{0}
\renewcommand\thesubfigure{(\alph{subfigure})}
\subfigure[Workload A.]{
    \label{fig:eval:ycsb:a}
    \includegraphics[width=0.205\textwidth]{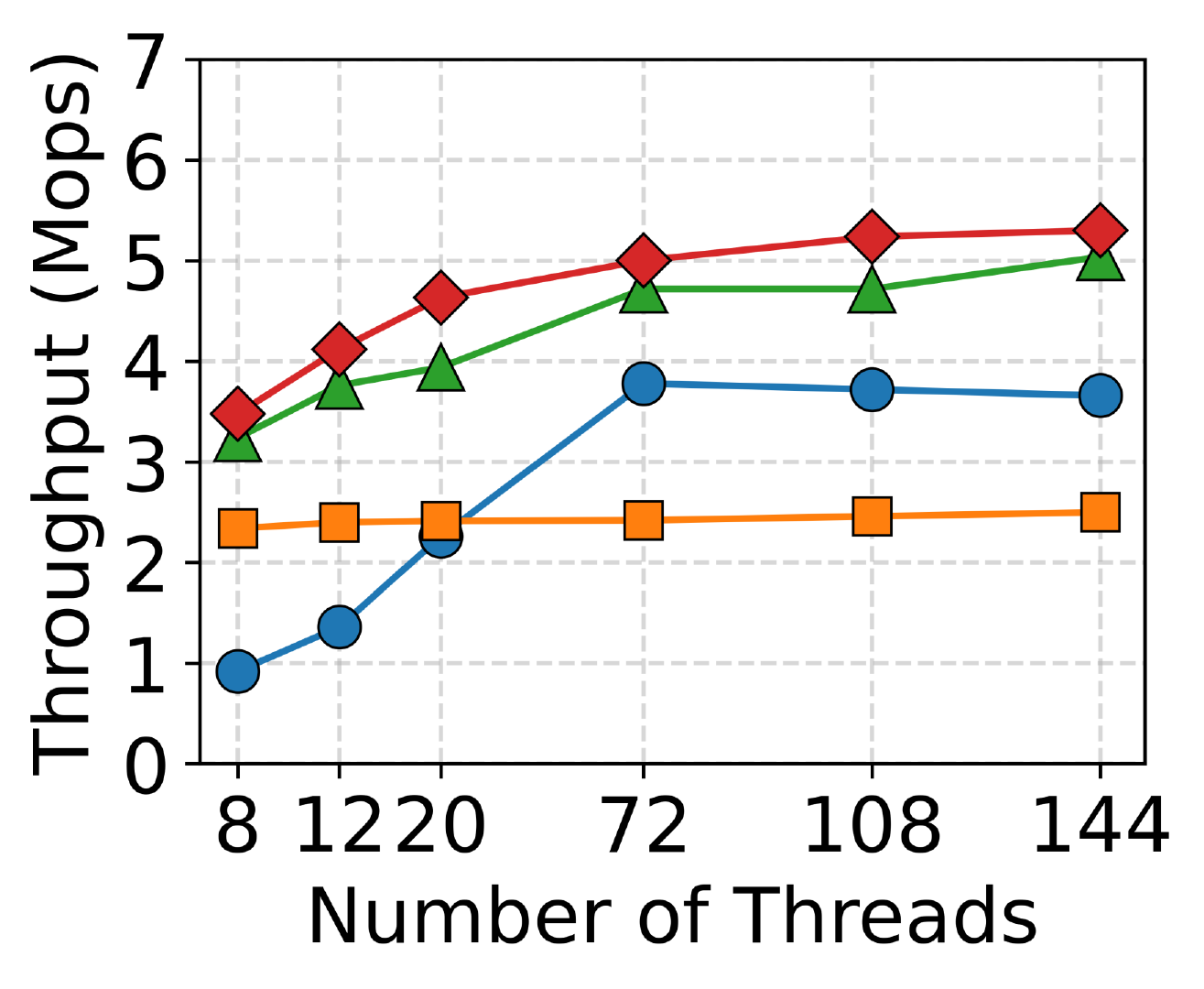}}
\hspace{-2.ex}
\subfigure[Workload B.]{
    \label{fig:eval:ycsb:b}
    \includegraphics[width=0.20\textwidth]{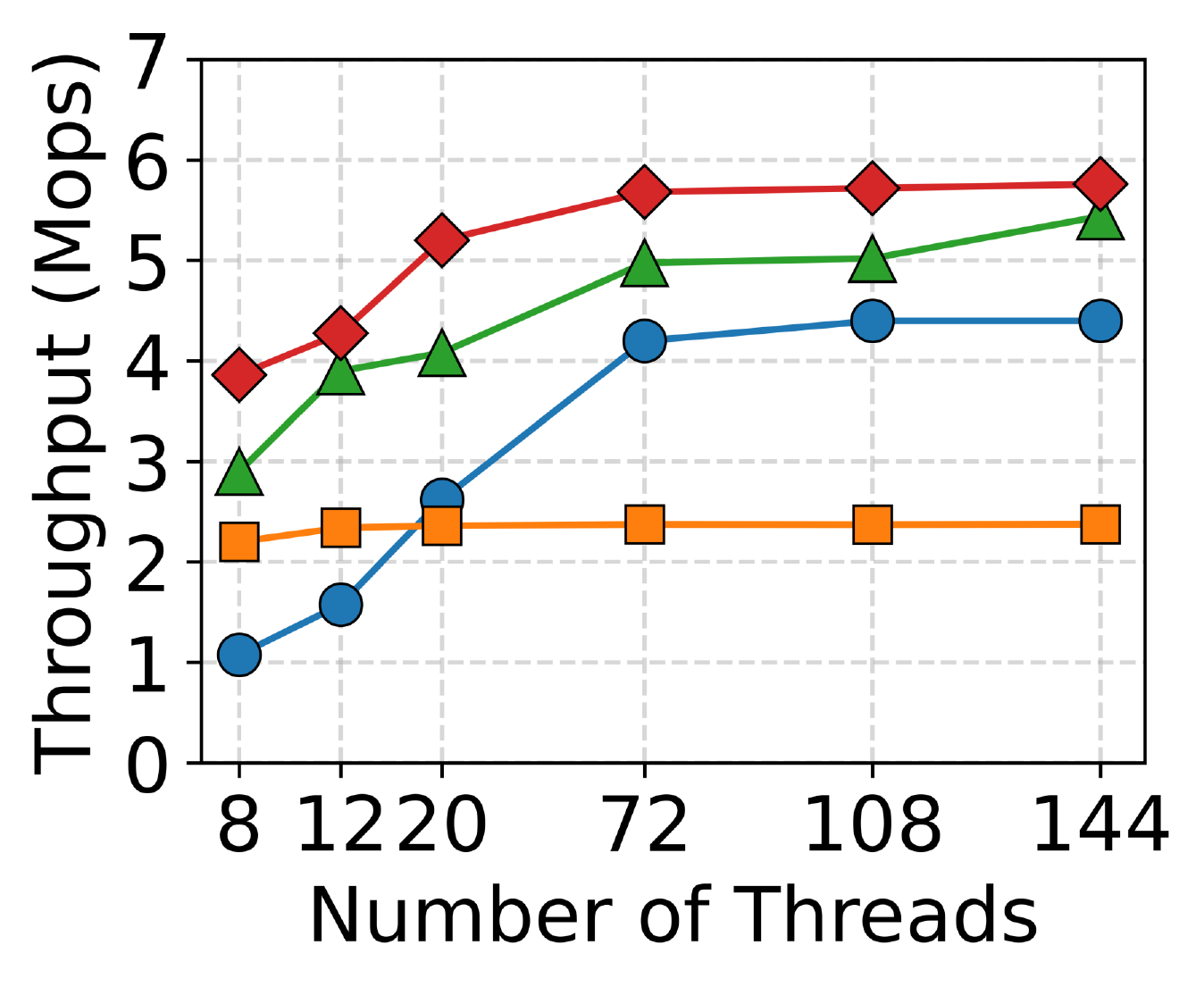}}
    \hspace{-2.5ex}
\subfigure[Workload C.]{
    \label{fig:eval:ycsb:c}
    \includegraphics[width=0.20\textwidth]{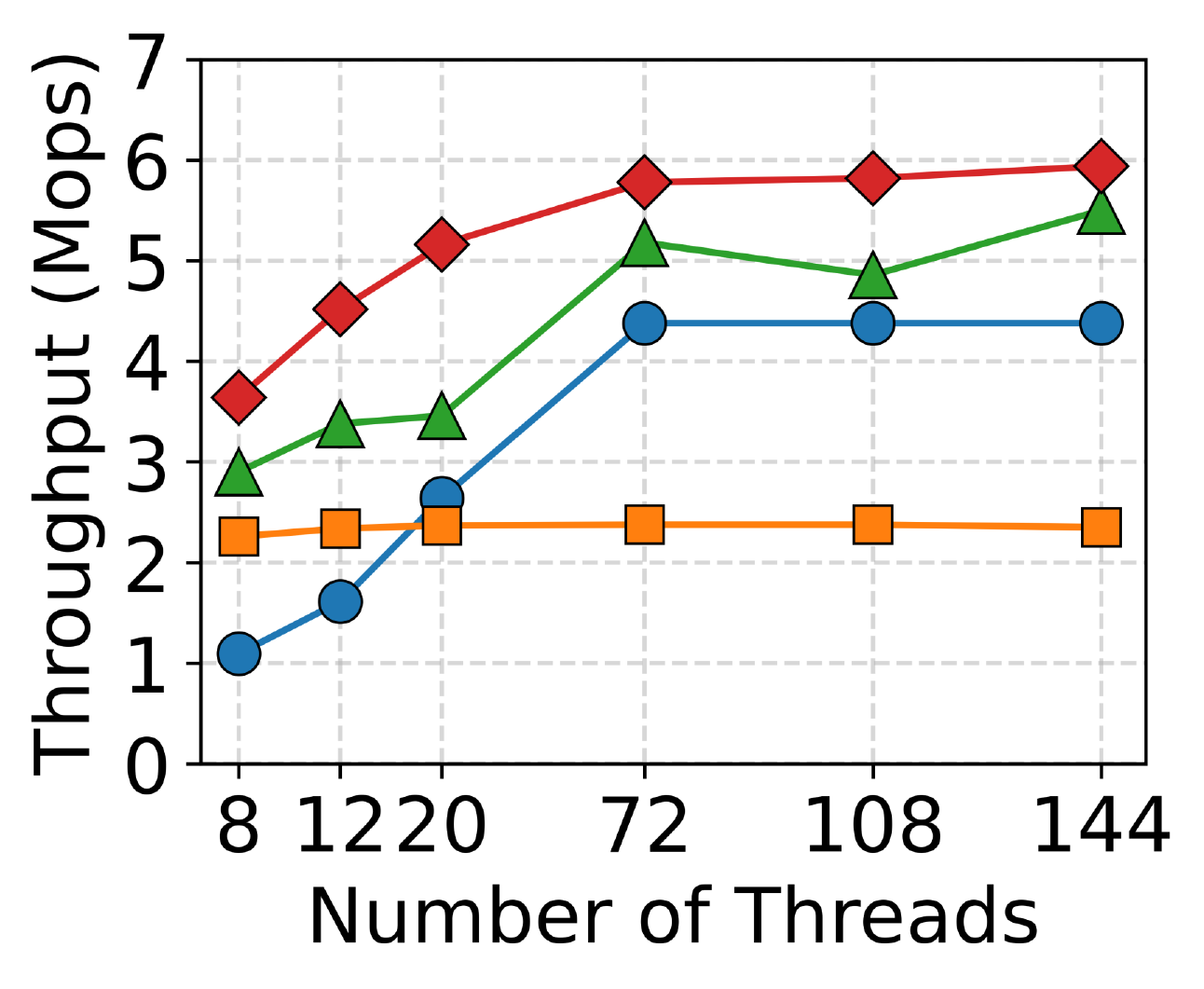}}
    \hspace{-2.5ex}
\subfigure[Workload D.]{
    \label{fig:eval:ycsb:d}
    \includegraphics[width=0.20\textwidth]{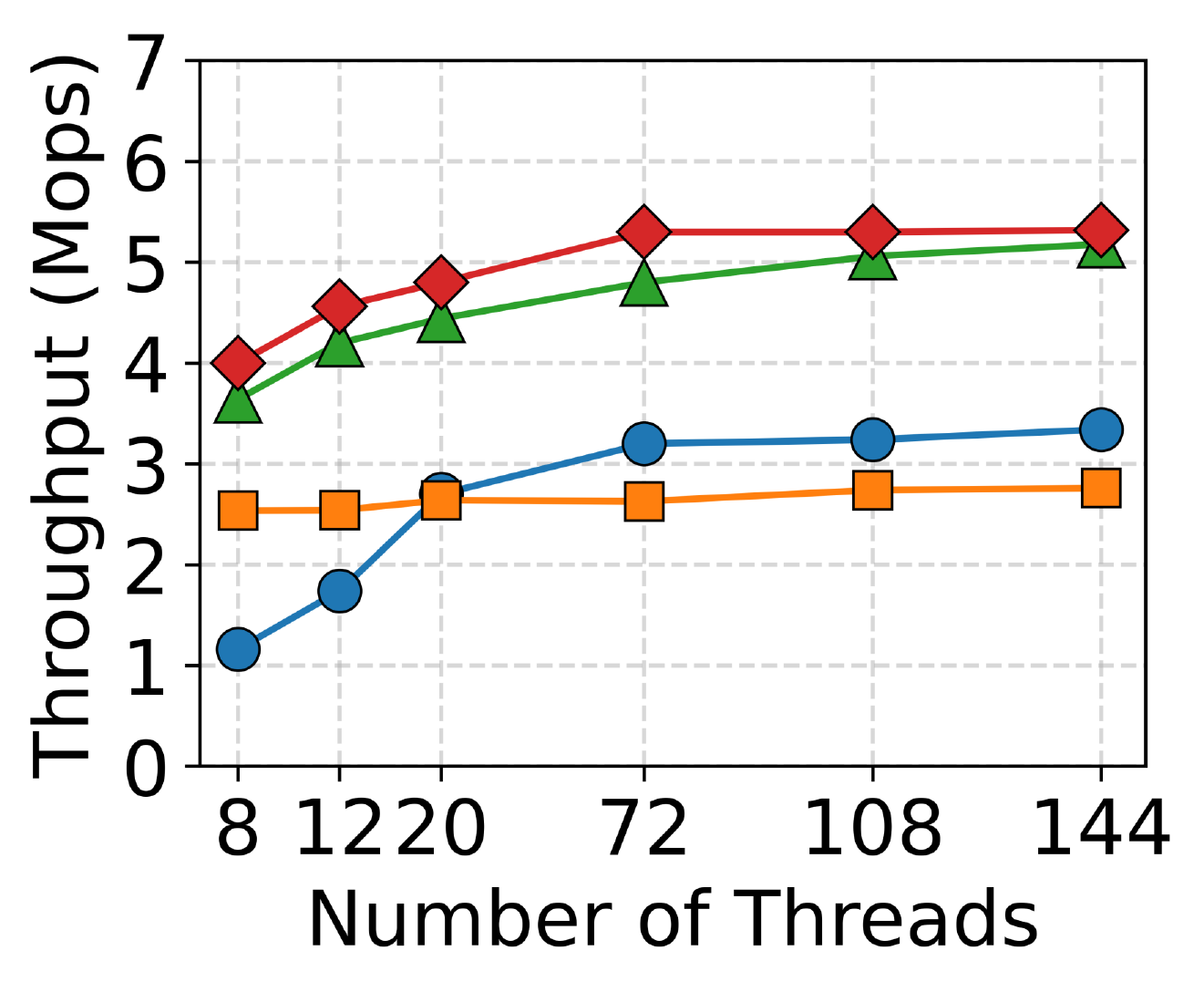}}
    \hspace{-2.5ex}
\subfigure[Workload F.]{
    \label{fig:eval:ycsb:f}
    \includegraphics[width=0.20\textwidth]{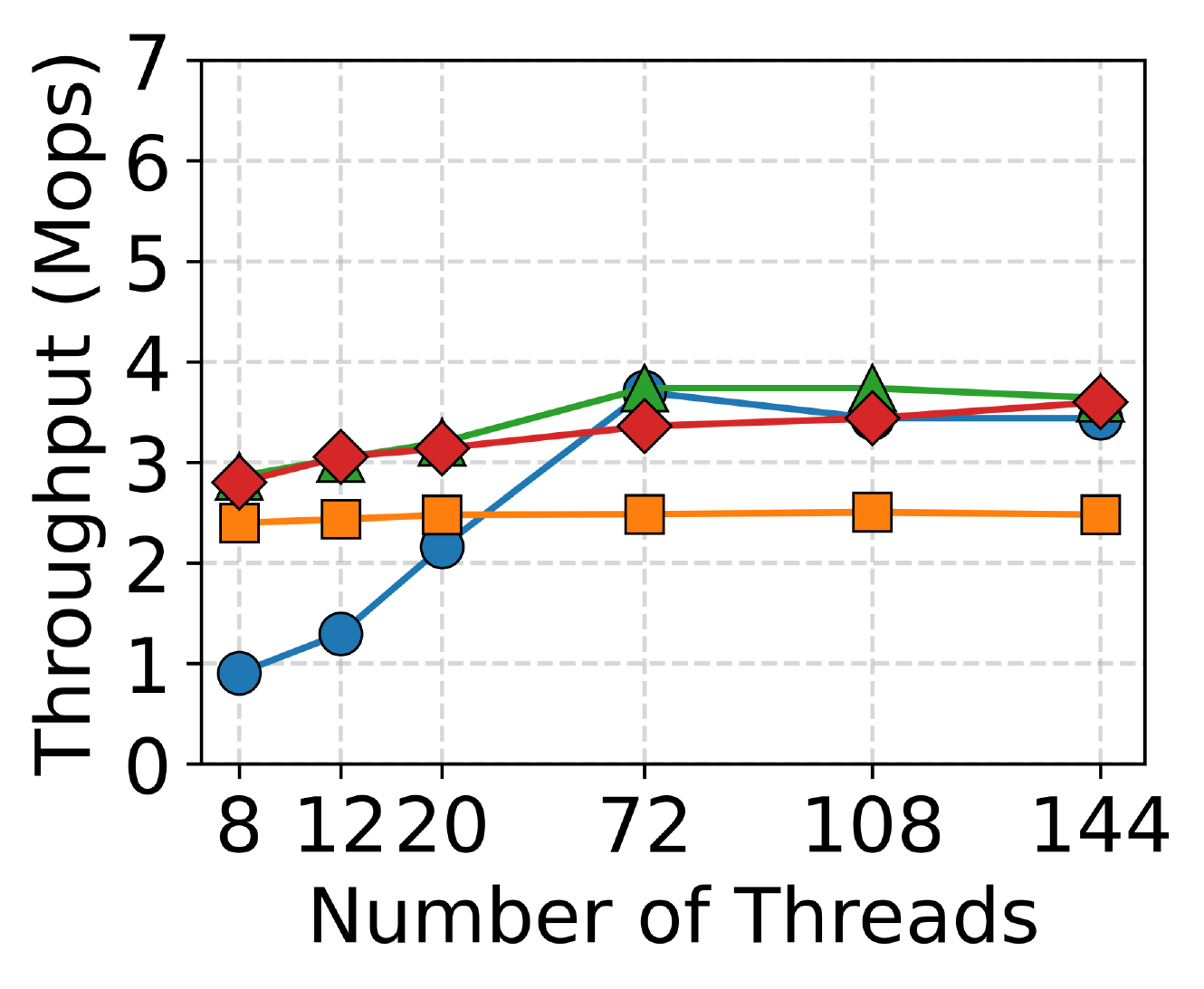}}
\vspace{-3ex}
\caption{\red{Throughput under YCSB benchmark with single memory node thread with Mellanox CX-6.}}
\vspace{-2.ex}
\label{fig:eval:ycsb}
\end{figure*}

\begin{figure*}[!t]
\centering
\renewcommand\thesubfigure{}
\subfigure[]{
    \includegraphics[width=0.58\textwidth]{Figures/legend.pdf}}\\
\vspace{-6ex}
\setcounter{subfigure}{0}
\renewcommand\thesubfigure{(\alph{subfigure})}
\hspace{-1.ex}
\subfigure[Workload A.]{
    \label{fig:eval:cx3:a}
    \includegraphics[width=0.202\textwidth]{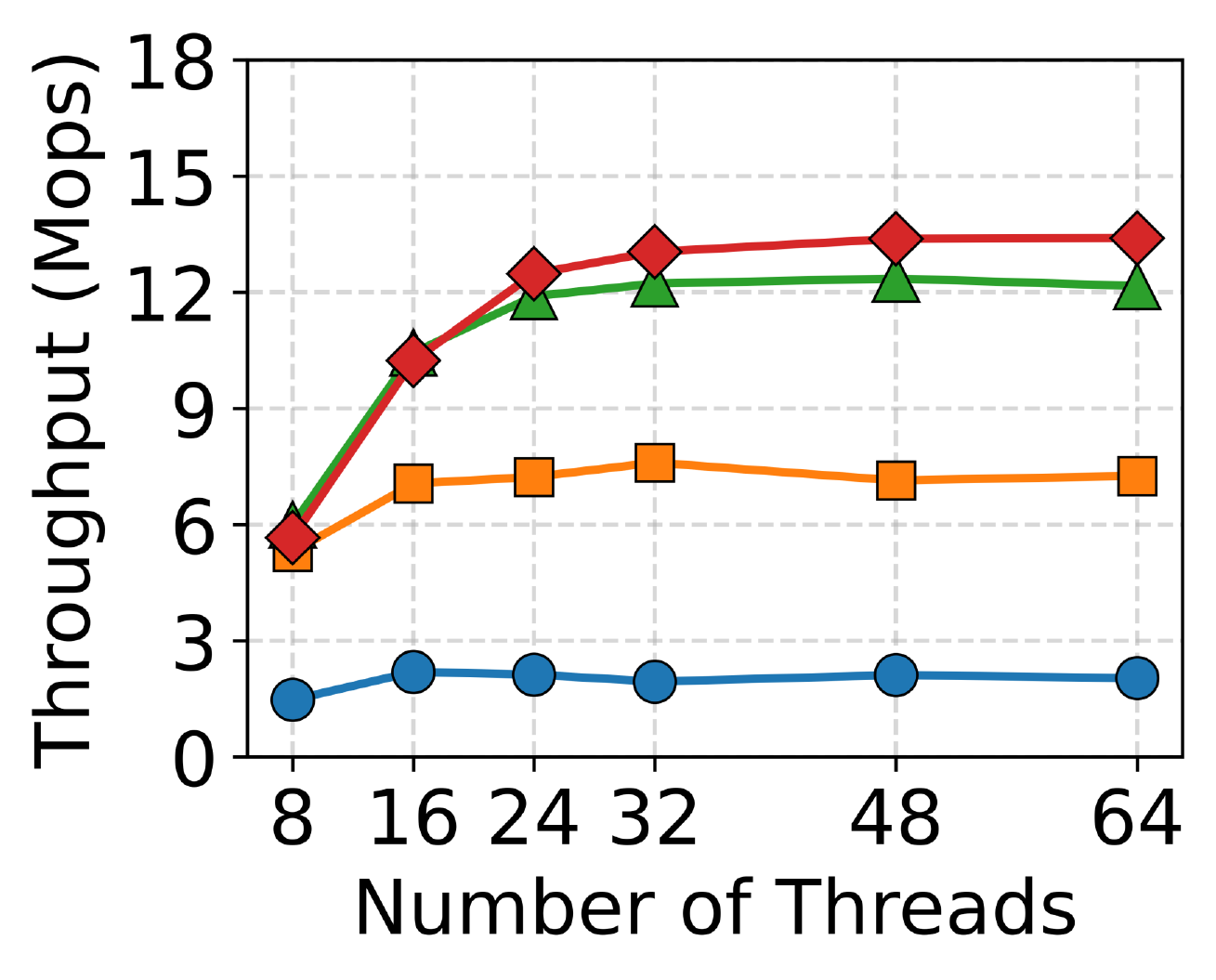}}
\hspace{-1.ex}
\subfigure[Workload B.]{
    \label{fig:eval:cx3:b}
    \includegraphics[width=0.20\textwidth]{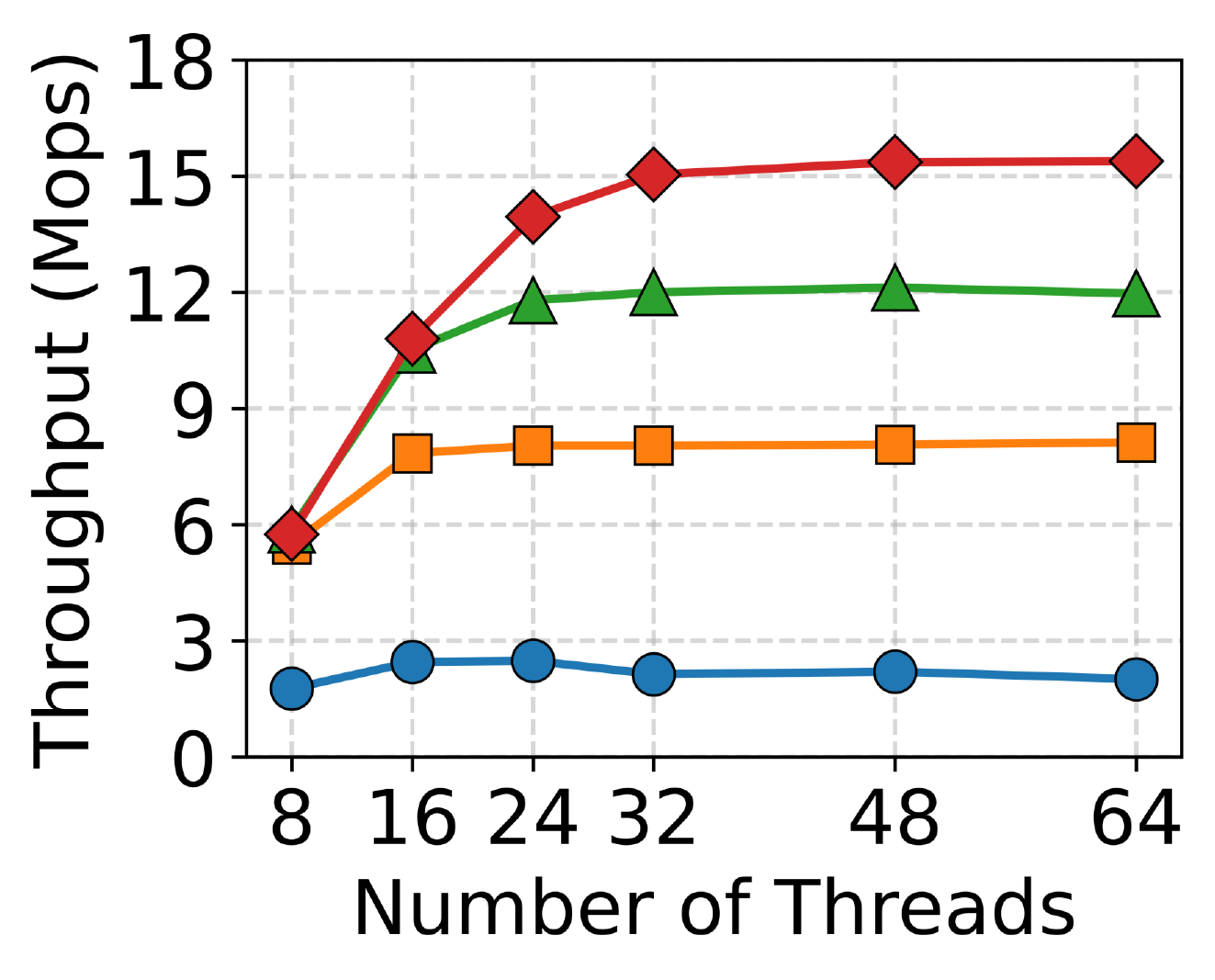}}
    \hspace{-2.ex}
\subfigure[Workload C.]{
    \label{fig:eval:cx3:c}
    \includegraphics[width=0.20\textwidth]{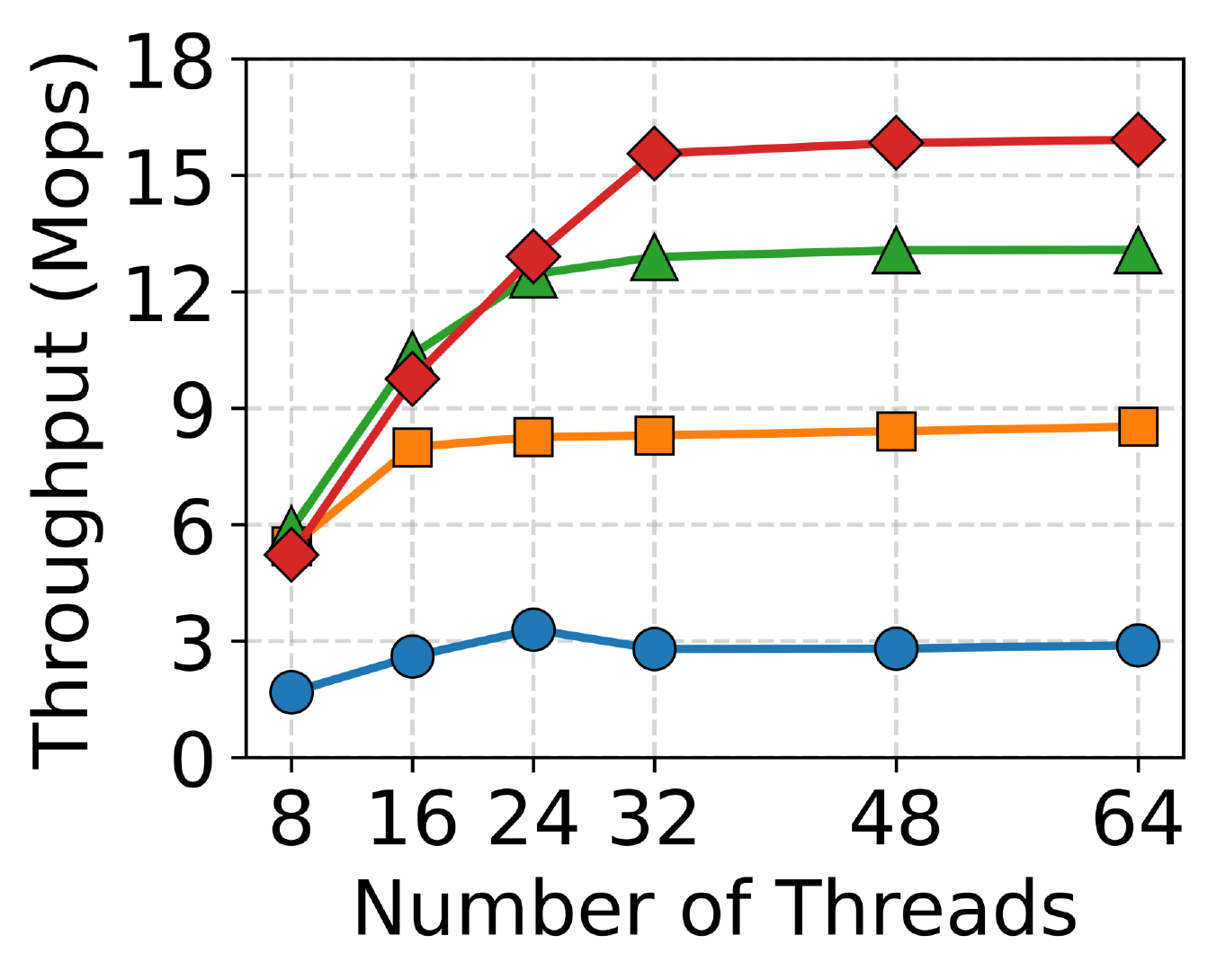}}
    \hspace{-2.ex}
\subfigure[Workload D.]{
    \label{fig:eval:cx3:d}
    \includegraphics[width=0.20\textwidth]{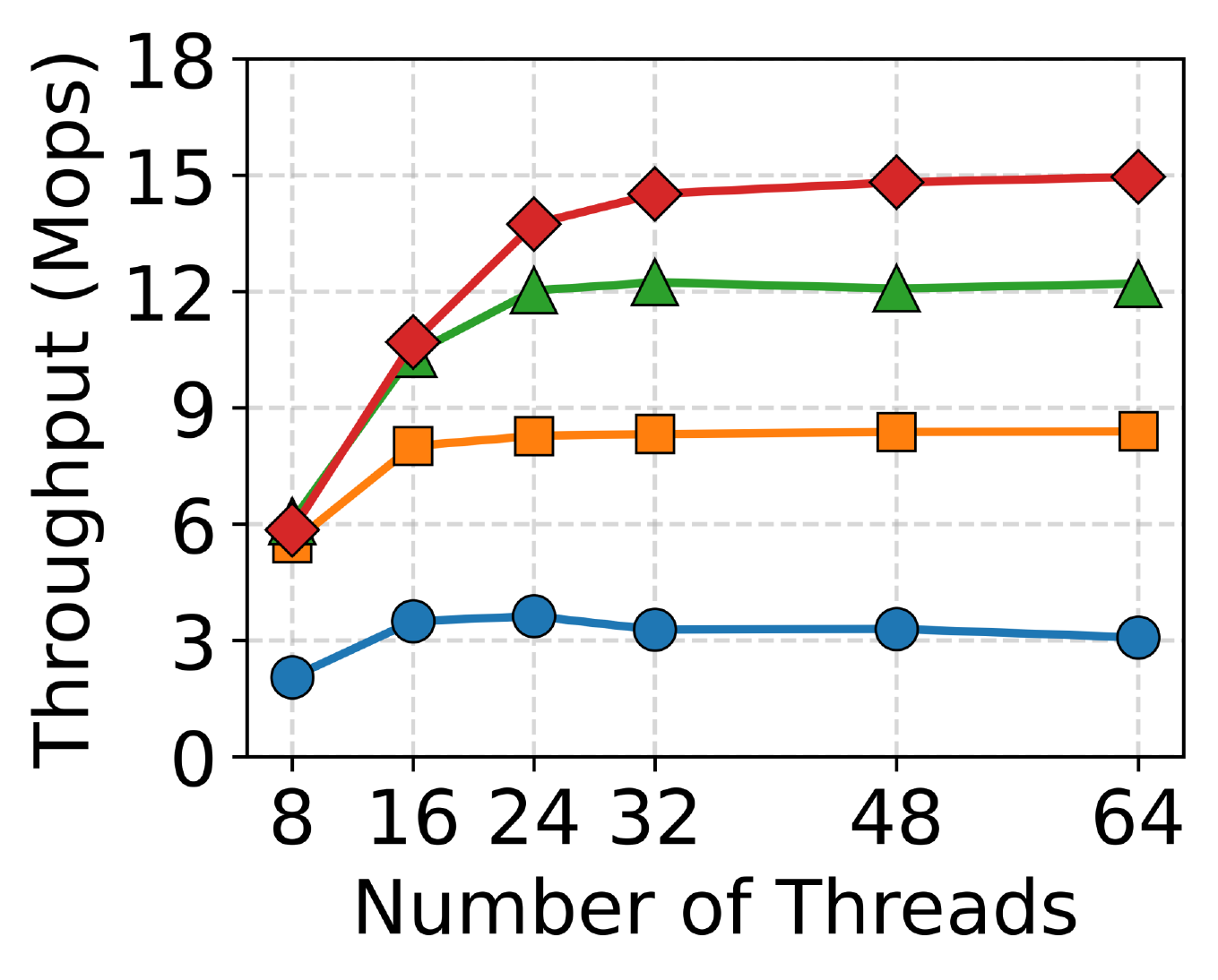}}
    \hspace{-2.ex}
\subfigure[Workload F.]{
    \label{fig:eval:cx3:f}
    \includegraphics[width=0.20\textwidth]{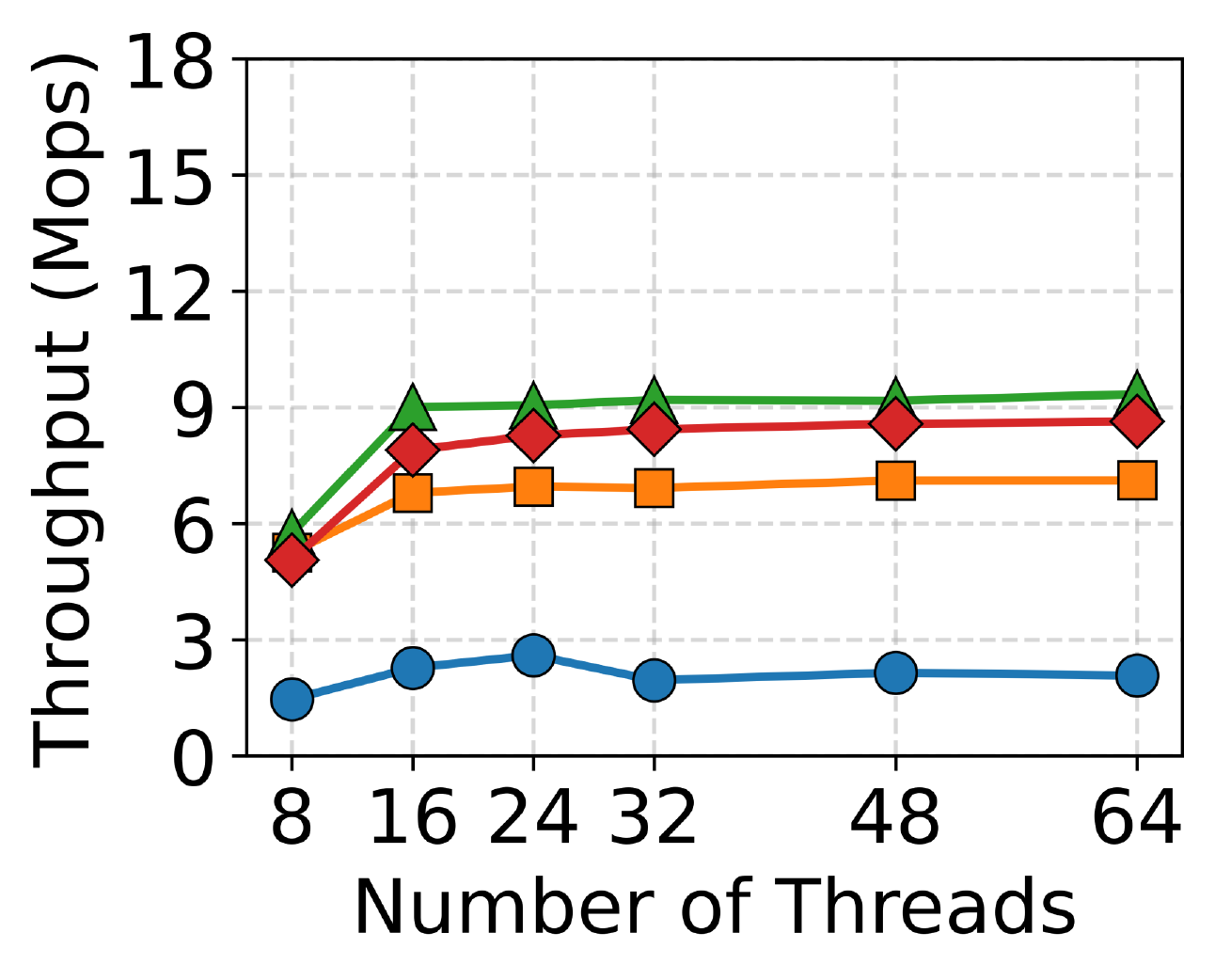}}
\vspace{-3ex}
\caption{Throughput under YCSB benchmark with Mellanox CX-3 RNICs.}
\vspace{-3.ex}
\label{fig:eval:cx3}
\end{figure*}

\subsection{Methodology}
\textbf{Testbed.}
\red{We run experiments in two environments. 1) 6 r650 machines from a public cluster CloudLab~\cite{cloudlab}; each of them is equipped with one Two 36-core Intel Xeon Platinum 8360Y CPU at 2.4GHz, 256 GiB DRAM and one Dual-port Mellanox ConnectX-6 (CX-6) 100 GbE NIC with Driver version as MLNX\_OFED\_LINUX-4.9-5.1.0.0. 
We conduct experiments with two shards, and each shard contains 3 machines. We use one machine as the memory node and the other two as compute nodes.
The memory node registers the memory with huge pages to reduce RNIC's page cache misses, which is beneficial for memory-intensive applications~\cite{xtore, race}. On compute nodes, we use two coroutines on each client thread to increase the query efficiency (See analysis in Section.~\ref{sec:eval:coros}).
This is the default experiment environment unless otherwise stated. 
2) 9 r320 machines in CloudLab~\cite{cloudlab}, each of them is equipped with one Xeon E5-2450 CPU (8 cores, 2.1Ghz), 16 GiB DRAM, and one Mellanox MX354A Dual port FDR CX3 adapter. We use 1 machine as the memory node and the other 8 as compute nodes.
We utilize 64-byte RDMA messages for all workloads to encapsulate various operation types (RC READ, UD SEND, and UD RECV), ensuring each request is padded to span two cache lines~\cite{guidelines}. We do not use batching at any layer to minimize the latency in all evaluations. %
}

\noindent\textbf{Workloads.} 
To evaluate the overall performance of \sys and other baselines, we employ YCSB~\cite{ycsb,ycsbc} workloads along with two diverse real-world datasets~\cite{sosd}. These datasets are 
(1) FB, encompassing a random assortment of Facebook user IDs to analyze patterns within social media interactions; (2) OSM, providing digitized infrastructure footprints from Open Street Map to represent geographical and spatial data usage; 
To ensure the datasets reflect general, unsorted data conditions, we shuffle them if initially sorted upon loading.
Unless specified, we use 8B keys and 8B address values to configure all workloads like existing schemes~\cite{rolex, learnedindex} for comprehensive evaluations. 
For each run, we precondition the memory node and warm up the database with 64 million KV pairs at first and then issue 10M requests to the benchmark on top of it.

\noindent\textbf{Baselines.}
We develop a prototype of \sys based on RDMA libraries rlib and r2~\cite{drtmh} with over 4000 LoC in C++. We compare \sys with the other three baselines, one is a recently proposed one-sided RDMA scheme, RACE hashing~\cite{race}, which utilizes RDMA RC READs for its operations; The other two are two-sided RDMA schemes that operate on RDMA SENDS/RECVs, differing in their underlying data structures -- MICA~\cite{herd, mica} and Cluster hashing~\cite{drtmr}.

\begin{itemize}[left=0em]
\vspace{-.3ex}
    \item \textbf{RACE hashing.}
    RACE hashing~\cite{race} is a representative one-sided RDMA scheme developed recently. It offloads all data operations to compute nodes to free the memory node CPU with one-sided RDMA primitives. RACE Hashing adopts an RDMA-friendly hash table to combine the overflow bucket for collided keys and the hashed bucket. Thus, all the candidate buckets containing the requested key can be read back together.
    We develop RACE hashing with over 1,400 lines of C++ code, excluding the benchmark part that is shared with other baselines.
    \item \textbf{RDMA RPC-MICA.}
    RPC-MICA is a two-sided RDMA-based scheme with a data structure MICA~\cite{mica,herd}, which is an efficient hopscotch hash table and it has been used in existing two-sided RDMA~\cite{herd,fasst}. The overflowed KV pairs can be stored in the bucket adjacent to its hashed bucket.
    We implement hash computation for the bucket number on the compute node and send the queried key's fingerprint and bucket number to save computation on the memory node. 
    We apply the open-source code from MICA~\cite{mica} in our benchmark, utilizing it as the underlying data structure for the RPC-based approach without batching.
    \item \textbf{RDMA RPC-Cluster hashing.}
    RPC-Cluster hashing is a two-sided RDMA baseline with Cluster hashing, a chained-based hash table with associativity, running on memory nodes~\cite{drtmh,drtmr}.
    The overflow keys that are hashed to a full bucket will be put in the linked indirect bucket. Each slot in a bucket includes 14 bits of fingerprint for key comparison. We apply the open-source code~\cite{drtm_code} of the cluster hashing as the data backend of our RPC-based scheme suit. 
    \vspace{-1ex}
\end{itemize}

\vspace{-1ex}
\subsection{Performance on YCSB}
\label{sec:eval:ycsb}


\red{\textbf{Performance with CX-6 RNICs.} We show the throughput of all evaluated methods by increasing the request load of running 8, 12, 20, 72, 108, and 144 compute node threads in a shard.
On the memory node, we consistently allocate only one thread to run on a single core. As shown in Fig~\ref{fig:eval:ycsb}, these five figures illustrate the throughput and latency results under YCSB workloads A, B, C, D, and F, respectively.}

\textbf{\texttt{Get} and \texttt{Update} workloads (YCSB A and B).}
YCSB A and B workloads include 50\% and 5\% data \texttt{Update} respectively and the remaining is \texttt{Get}. 
\sys can achieve 5.50 and 5.82 Mops throughput for YCSB A and B, as shown in Fig.~\ref{fig:eval:ycsb:a} and Fig.~\ref{fig:eval:ycsb:b}. 
All other methods show lower throughput with the same number of threads. 
\sys can provide up to 1.07$\times$ and 1.06$\times$ throughput improvements on workloads A and B respectively, compared to RPC-cluster hashing. 
Compared to other RPC baselines with associative hash tables, the memory node in \sys is offloaded with less computation because it only needs to read the targeted key, and no data probing or traversing is needed to find the targeted value of the key. 
RACE hashing requires three round trips for updating data consistently, significantly increasing the latency and limiting the throughput.
By comparing the results between workloads A and B, 
when more \texttt{Update} requests are issued, \sys spends more computation resources for value rewriting and key checking by reading the underlying KV blocks indicated by the computed MPH slot. 
Hence, \sys under YCSB B provides higher throughput than \sys under YCSB A.

\textbf{\texttt{Get}-only workload (YCSB C).}
For \texttt{Get}-only workload, \sys can achieve 6.01 Mops throughput.
When the number of compute node threads reaches 72, \sys outperforms RACE hashing, MICA, and Cluster hashing by 1.31$\times$, 2.43$\times$, and 1.11$\times$ on total throughput, respectively. The performance of RACE hashing is bottle-necked by its two round trips and the limited RNIC memory to cache queue pair (QP) state of a larger number of reliable connections.
\sys reduces the average memory node's CPU time for data \texttt{Get} request with less computation overhead than the other two RPC-based baselines while looking up a key.

\textbf{\texttt{Get} and \texttt{Insert} workloads (YCSB D and F).}
YCSB D contains  5\% \texttt{Insert} and 95\% \texttt{Get} operations. 
 YCSB F contains 25\% \texttt{Insert}, 25\% \texttt{Update}, and 50\% \texttt{Get} operations. 
Under YCSB D, \sys still shows the highest throughput among all methods. 
For \texttt{Insert} operations, \sys will check if a slot in the target bucket is available. Key-checking is also required, and a new seed will be calculated if the target slot stores an existing value. 
The high rate of \texttt{Insert} operations in YCSB F pulls the throughput down to 3.62 Mops, which is similar to RPC-Clustering hashing (3.64 Mops) when the number of client threads reaches 144.

\red{\textbf{Performance with CX-3 RNICs.}
As shown in Fig.~\ref{fig:eval:cx3}, we show the throughput with the 4 memory node threads and a set of compute node threads numbers 8, 16, 24, 32, 48, and 64, respectively. 
\sys can consistently achieve the highest throughput for read-intensive workloads (A, B, C, and D). 
Significantly, \sys outperforms RACE hashing, MICA, and Cluster hashing by 5.03$\times$, 1.79$\times$, and 1.23$\times$ on total throughput for workload C, respectively. 
When we use a weaker CPU, the advantage of \sys is more significant. Unfortunately, CloudLab does not offer a weaker CPU with a high-performance network. }

In summary, \sys demonstrates the highest throughput for most types of workload (YCSB A, B, C, and D). For a workload that is \texttt{Insert}-intensive such as YCSB F, \sys provides comparable throughput to other RDMA-RPC methods but still higher than that of one-sided RDMA.

\vspace{-2ex}
\subsection{\red{Evaluations on Real-World Datasets}}
\label{sec:eval:sosd}

\begin{figure}[!t]
\centering
\renewcommand\thesubfigure{}
\subfigure[]{
    \includegraphics[width=0.48\textwidth]{Figures/legend.pdf}}\\
\vspace{-5ex}
\setcounter{subfigure}{0}
\renewcommand\thesubfigure{(\alph{subfigure})}
\subfigure[Dataset FB, uniform workload.]{
    \label{fig:eval:sosd:a}
    \includegraphics[width=0.23\textwidth]{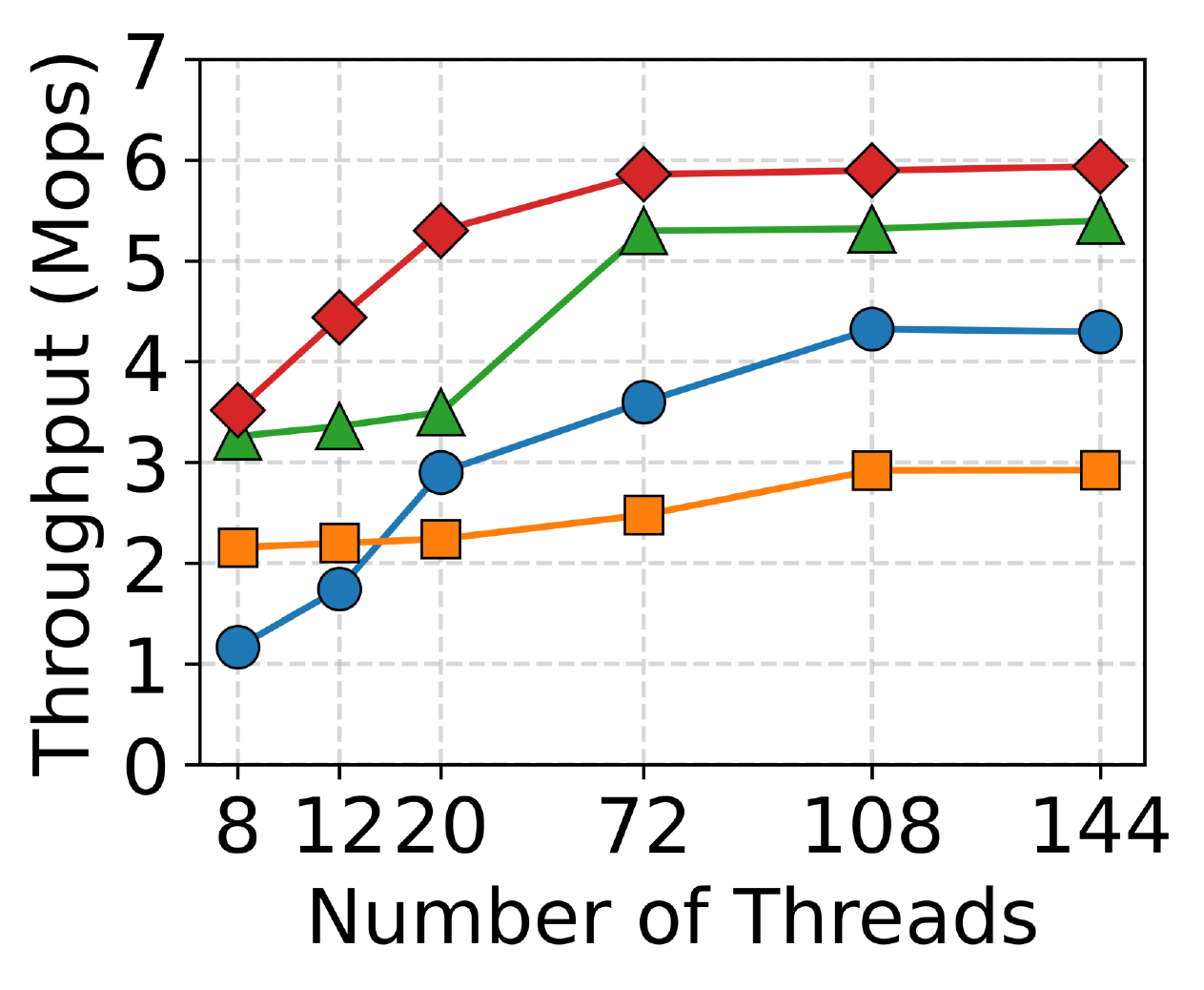}}
\subfigure[Dataset OSM, uniform workload.]{
    \label{fig:eval:sosd:b}
    \includegraphics[width=0.23\textwidth]{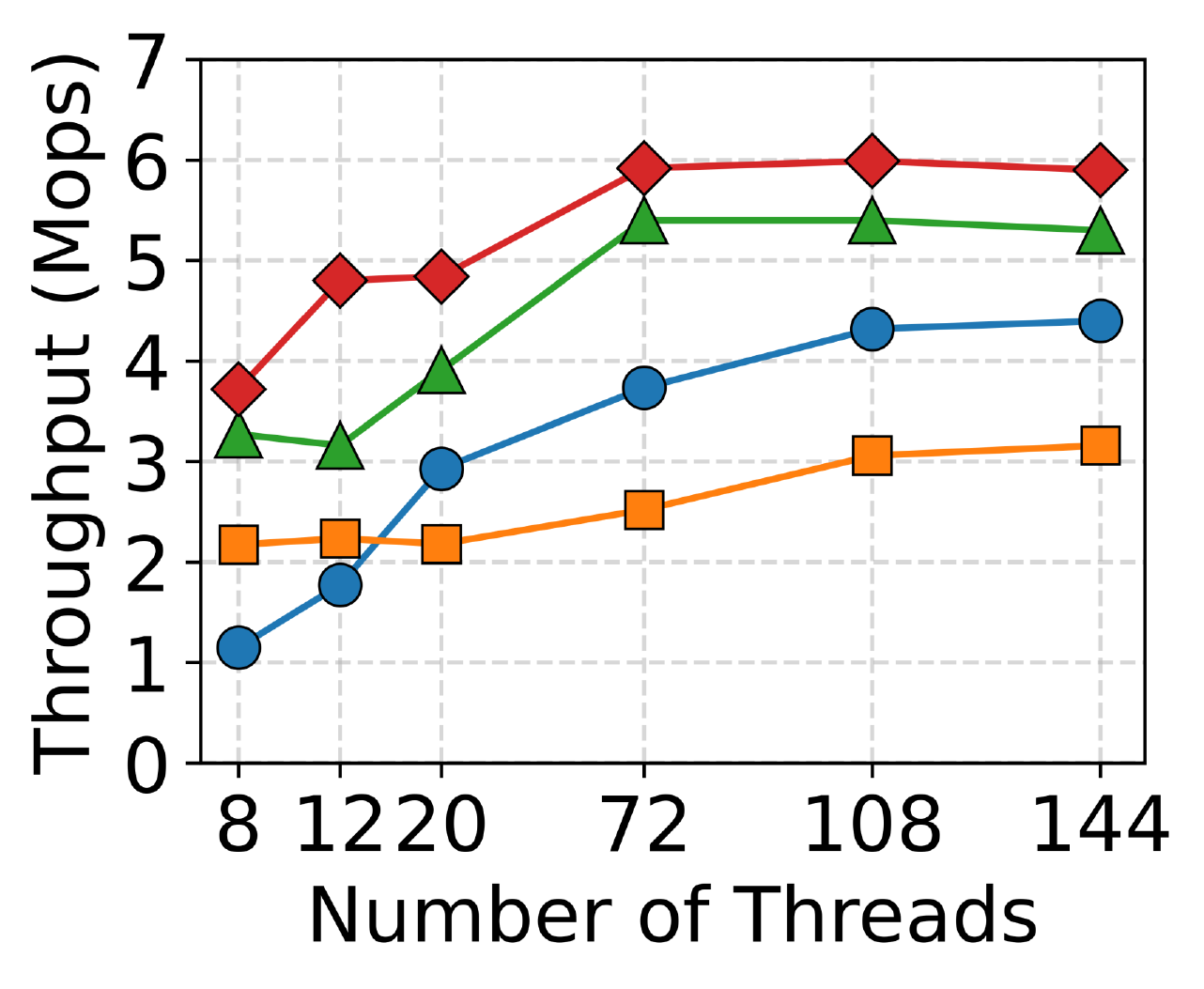}}\\
\vspace{-2.5ex}
\subfigure[Dataset FB, zipfian workload.]{
    \label{fig:eval:sosd:c}
    \includegraphics[width=0.23\textwidth]{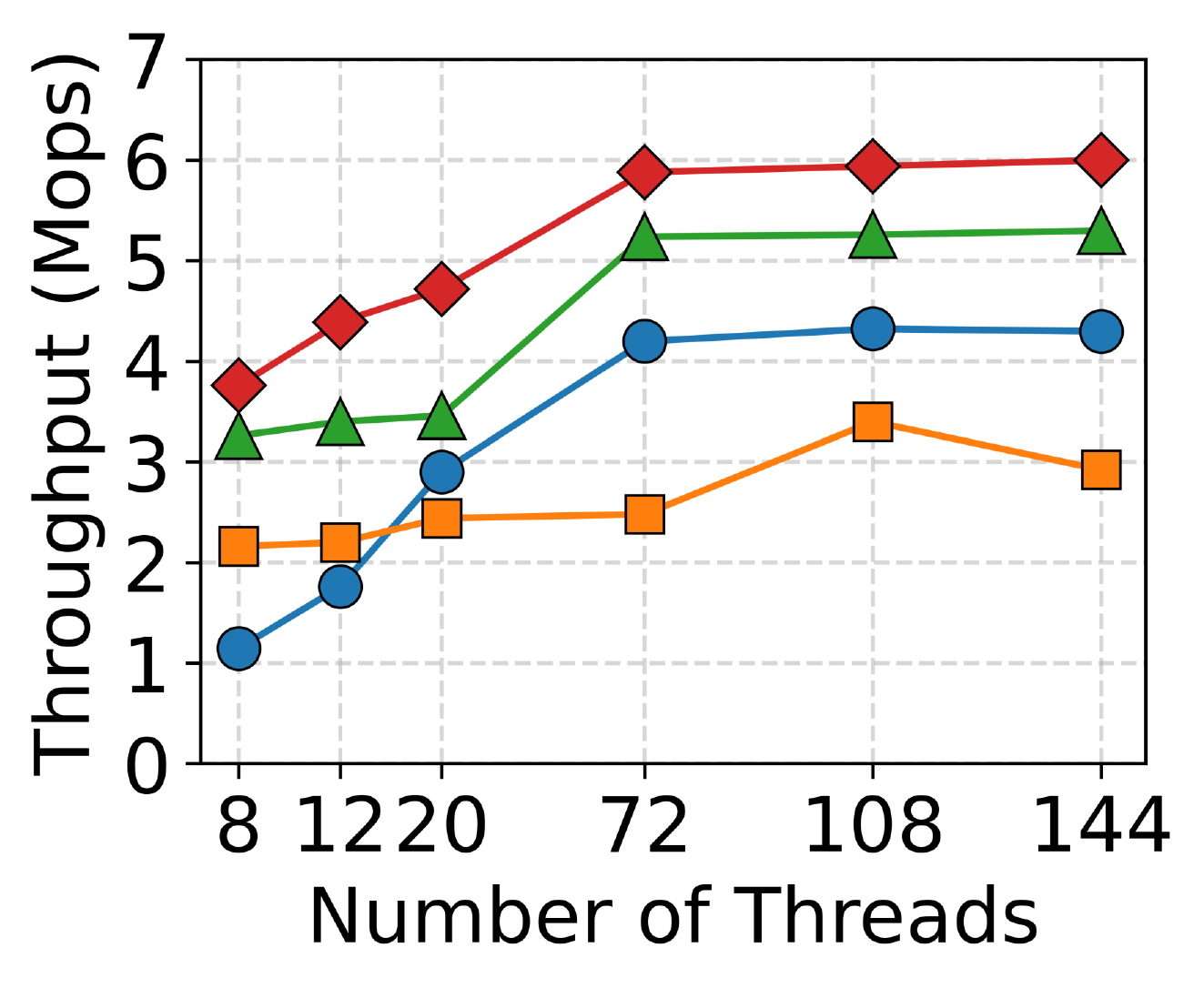}}
\subfigure[Dataset OSM, uniform workload.]{
    \label{fig:eval:sosd:d}
    \includegraphics[width=0.23\textwidth]{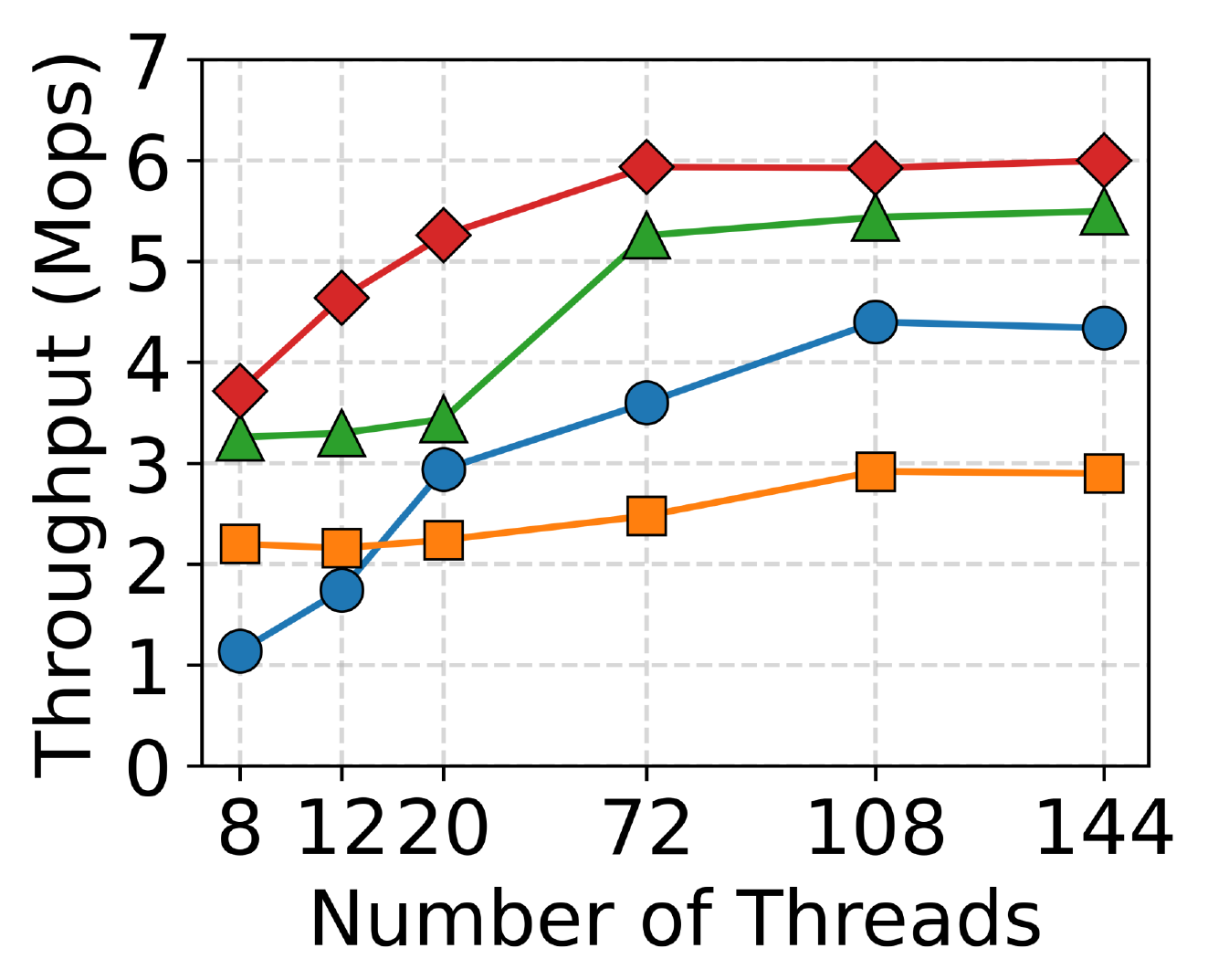}}
\vspace{-2.5ex}
\caption{\red{Data \texttt{Get} throughput performance with SOSD datasets with uniform and zipfian-0.99 workloads.}}
\label{fig:eval:sosd}
\vspace{-3.5ex}
\end{figure}

\red{We leverage the SOSD datasets~\cite{sosd} for evaluations. Fig.~\ref{fig:eval:sosd} illustrates throughput results with the number of compute node threads as 8, 12, 20, 72, 108, and 144 in a shard. We set the number of memory node threads to 1. Each compute node thread issues 10 million key lookup requests selected from the datasets in a uniform or zipfian distribution.} 

\red{Compared to RACE, \sys achieves throughput of 1.38$\times$, 1.35$\times$, 1.39$\times$, and 1.38$\times$ respectively on these four different settings when the number of threads reaches 144.
RACE's performance is constrained by the multiple round trips.
Compared to RPC-MICA and Cluster hashing, \sys achieves a throughput of 2.03$\times$ and 1.1$\times$ respectively on dataset FB when the threads number reaches 144 in Fig.~\ref{fig:eval:sosd:a}.
The reason that \sys can outperform them is that \sys can go directly to access data without extra check computation and indirect data accessing to probe the hash chain or buckets.
Also, \sys outperforms RACE hashing, RPC-MICA and RPC-CLuster hashing by 1.35$\times$, 2.05$\times$, and 1.13$\times$ respectively on dataset FB when the workload follows the Zipfian distribution, as shown in Fig.~\ref{fig:eval:sosd:c}. We observe the same trend in performance comparison with the dataset OSM.}



\vspace{-2ex}
\subsection{\red{Scalability with memory node threads}}
\label{sec:eval:cores}

\red{In this set of experiments, we vary the number of memory node threads from 1 to 3 and observe the throughput of different methods using real-world datasets FB and OSM. 
To exhaust the CPU resources on the memory node side, we use four r650 servers as compute nodes with 288 compute node threads.}

\begin{figure}[!t]
\centering
\renewcommand\thesubfigure{}
\hspace{1.1ex}
\subfigure[]{
    \includegraphics[width=0.32\textwidth]{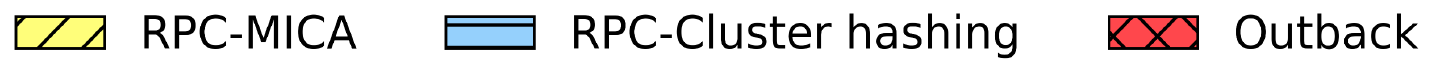}}\\
\vspace{-5.5ex}
\setcounter{subfigure}{0}
\renewcommand\thesubfigure{(\alph{subfigure})}
\subfigure[Scalability with memory node threads on dataset FB.]{
    \label{fig:eval:cores:a}
    \includegraphics[width=0.232\textwidth]{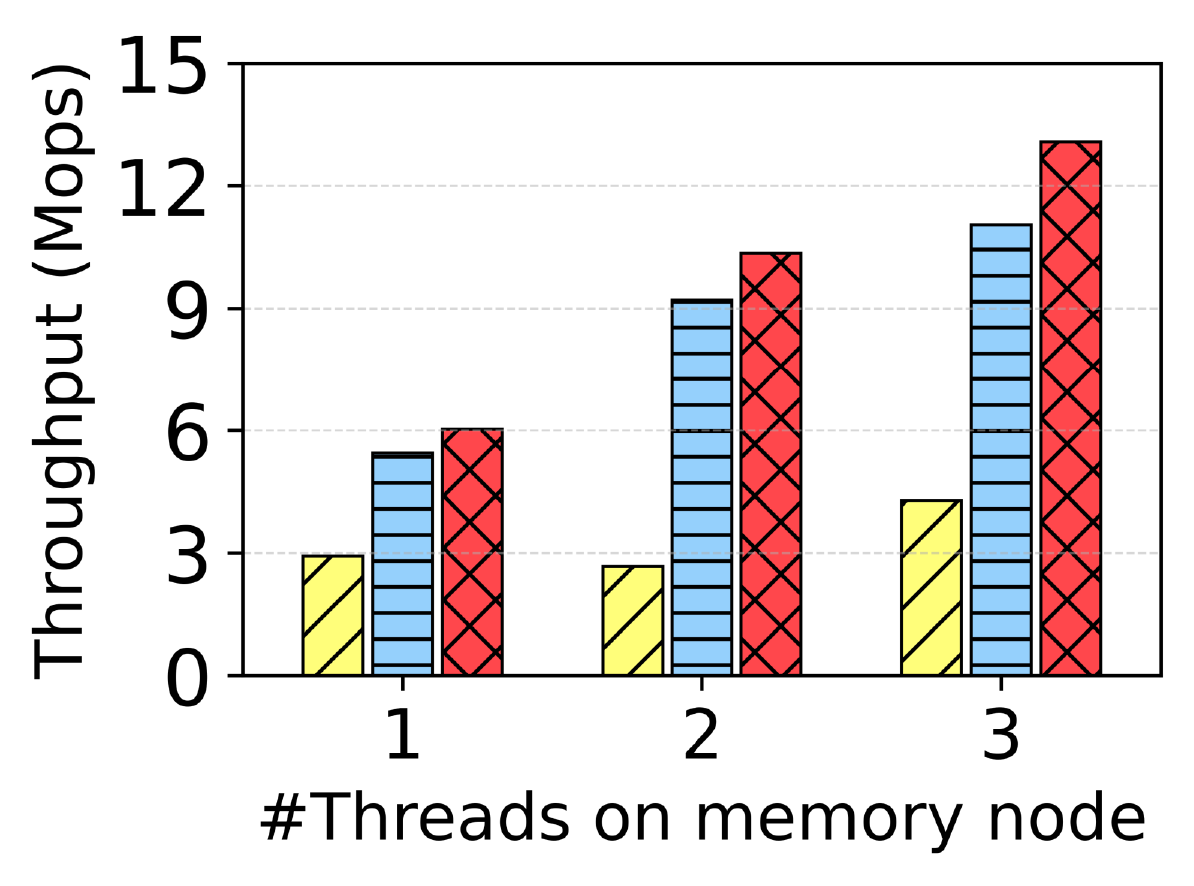}}
\hspace{-1.2ex}
\subfigure[Scalability with memory node threads on dataset OSM.]{
    \label{fig:eval:cores:b}
    \includegraphics[width=0.232\textwidth]{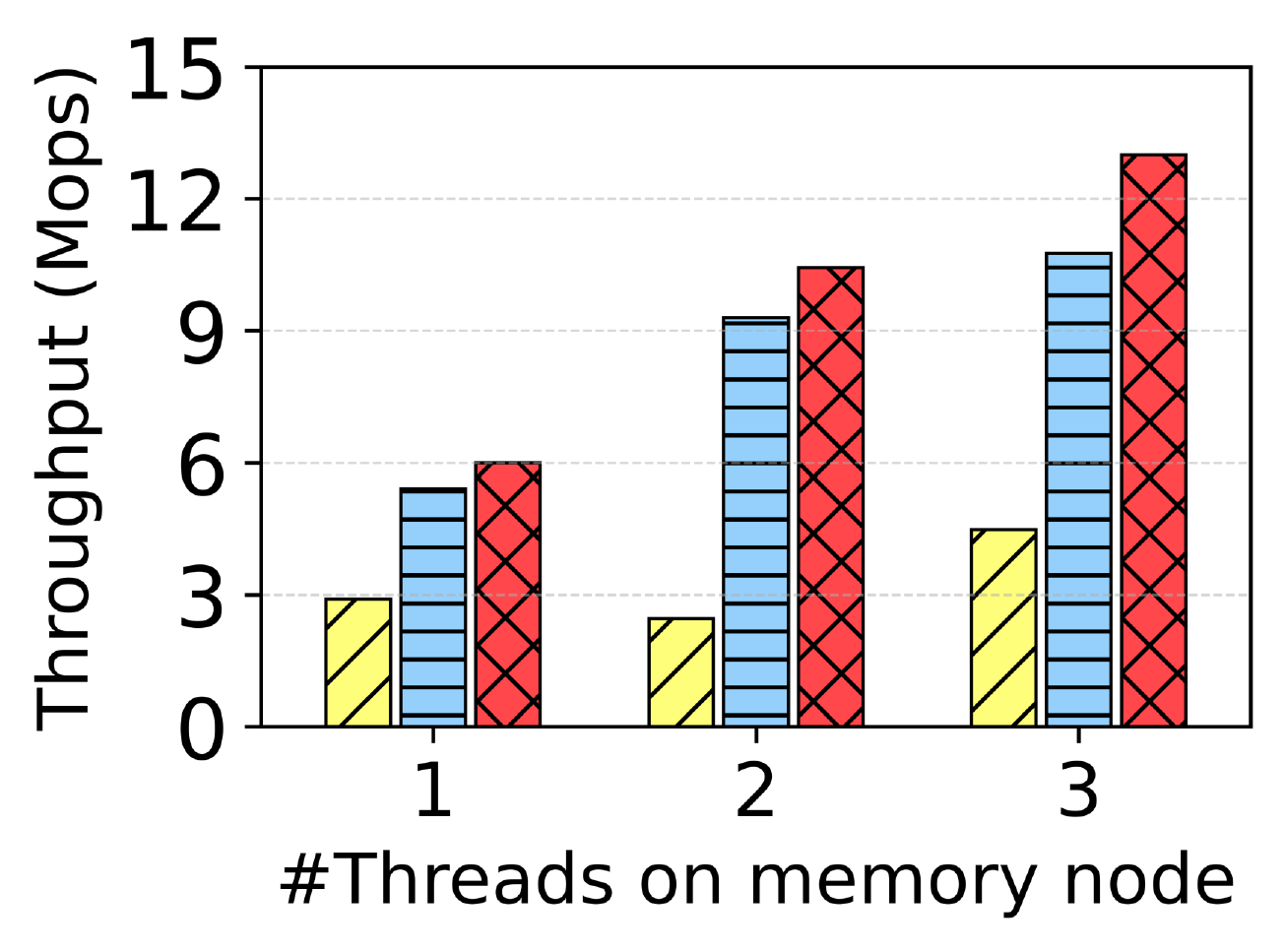}}
\vspace{-2.5ex}
\caption{\red{Throughput vs. the number of memory node threads.}}
\vspace{-3.5ex}
\label{fig:eval:cores}
\end{figure}

\red{Fig.~\ref{fig:eval:cores} shows the throughput of three RDMA-RPC schemes, by varying the memory node threads from 1 to 3. 
The throughput of \sys is around 1.10-1.21$\times$ of Cluster hashing and around 3$\times$ of MICA for dataset FB. 
The results of the two datasets exhibit the fact that as the number of compute node threads increases, the performance ratio between \sys and RPC-Cluster hashing/MICA remains similar.
The reason is that \sys can ease the CPU burden on the memory node and allow it to handle more data requests from the compute node threads by offloading the computation of indexing to compute nodes.}

\red{The fact that \sys achieves higher relative throughput to other RPC methods under a small number of memory node threads actually demonstrates the main advantage of \sys: achieving high performance when the memory node carries weak CPU power in a disaggregated memory system.}

\textbf{\red{Note that the aim of \sys is not to saturate RNIC but to increase the throughput when there are limited CPU resources in a memory node with two-sided RDMA primitives. 
The results in this section show that \sys can achieve higher CPU efficiency with the same throughput goal, and \sys can realize higher throughput with the same CPU resources. }
In disaggregated systems, this can motivate the industry to satisfy the user's throughput goal with less TCO by reducing the CPU resources equipped on memory-optimized cloud instances~\cite{ec2}.}

\vspace{-1.5ex}
\subsection{\red{Influence of the number of coroutines}}
\label{sec:eval:coros}

\begin{figure}[!t]
\centering
\renewcommand\thesubfigure{}
\hspace{1.1ex}
\subfigure[]{
    \includegraphics[width=0.42\textwidth]{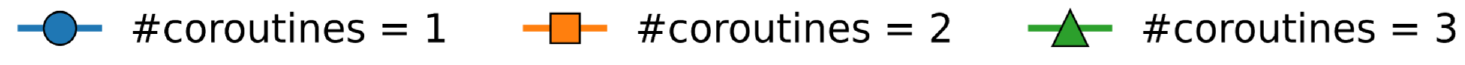}}\\
\vspace{-5.5ex}
\setcounter{subfigure}{0}
\renewcommand\thesubfigure{(\alph{subfigure})}
\subfigure[Latency-throughput curve on YCSB-C with 1 memory node thread.]{
    \label{fig:eval:coros:a}
    \includegraphics[width=0.235\textwidth]{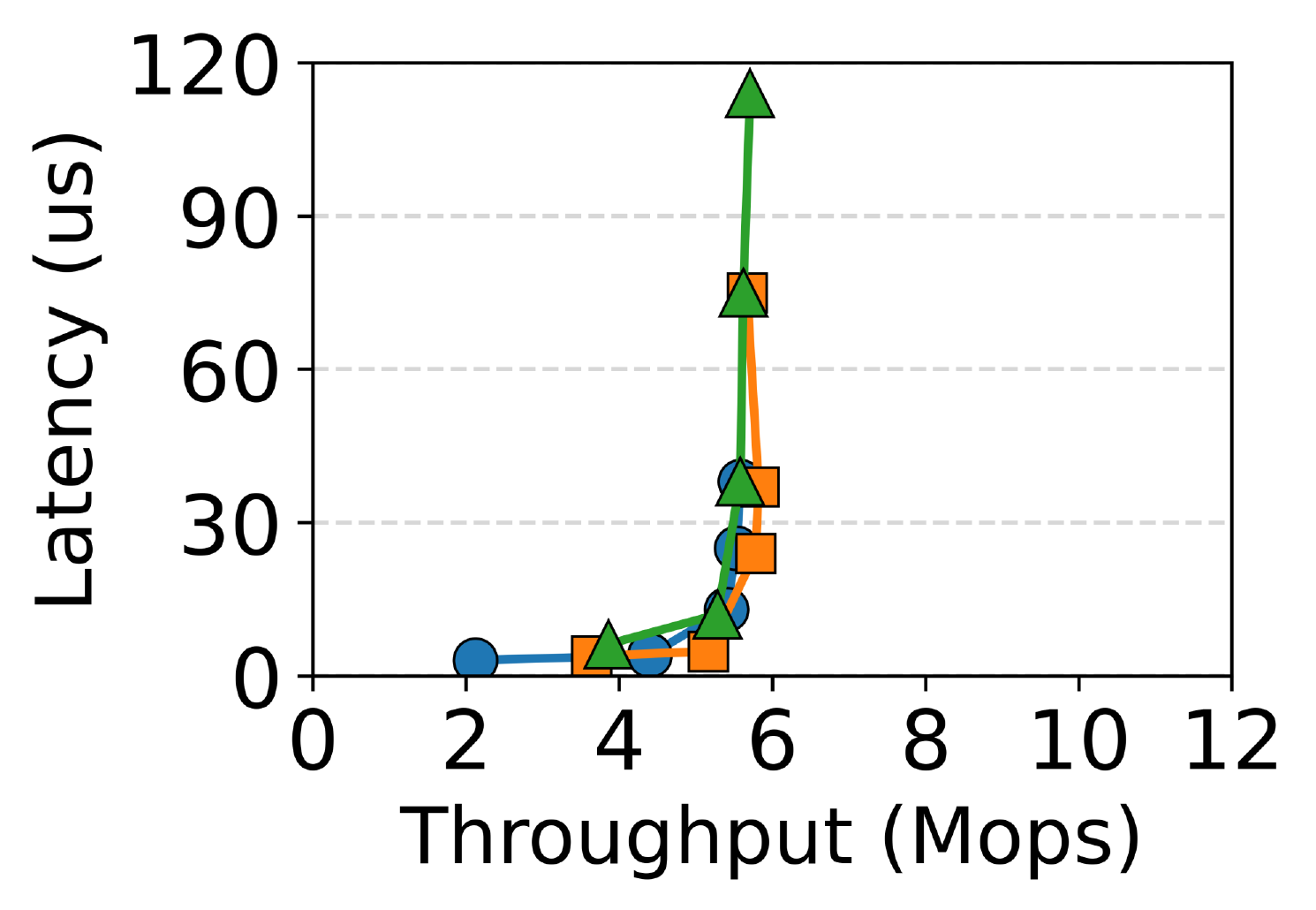}}
\hspace{-1.5ex}
\subfigure[Latency-throughput curve on YCSB-C with 2 memory node threads.]{
    \label{fig:eval:coros:b}
    \includegraphics[width=0.235\textwidth]{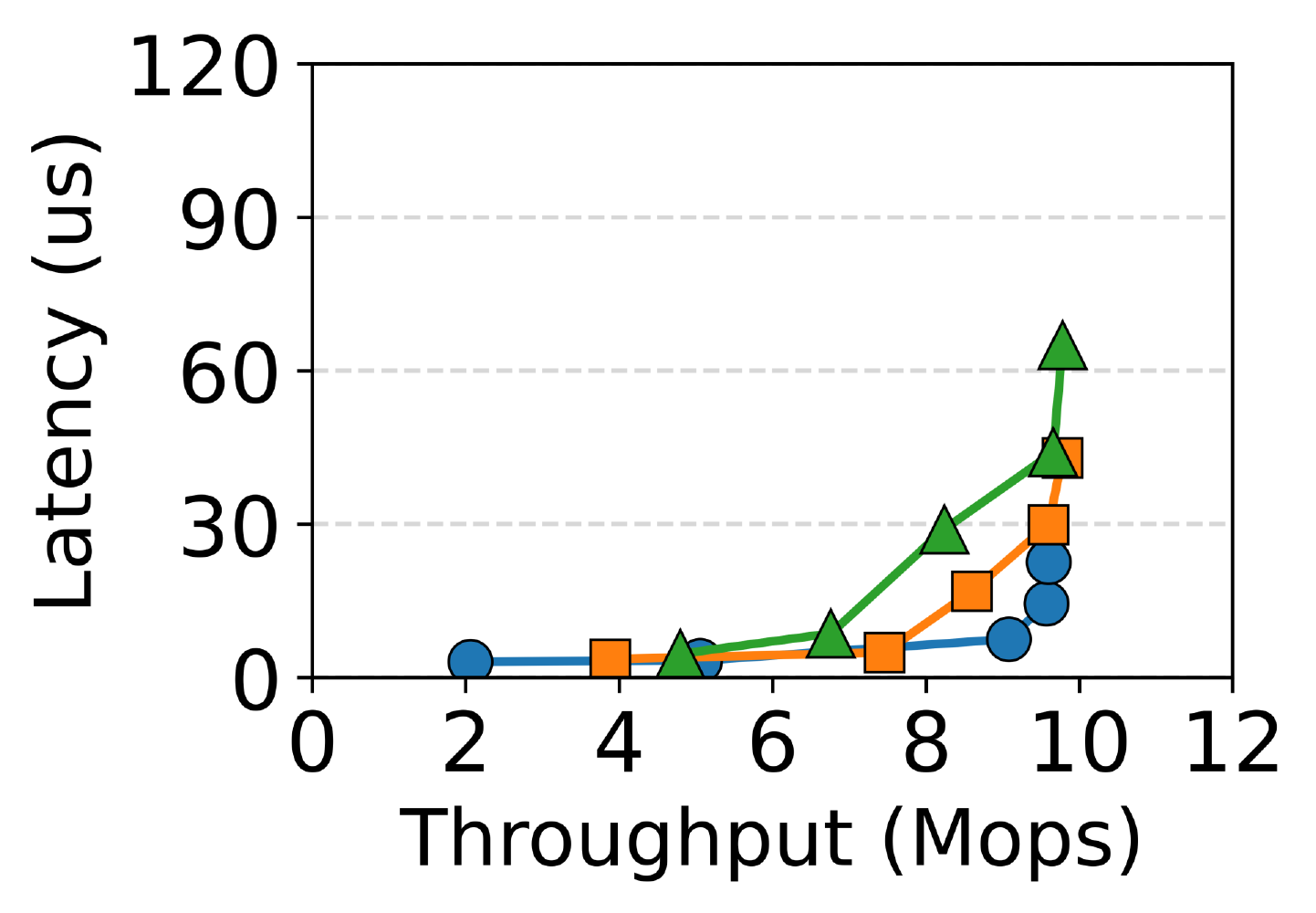}}
\vspace{-3ex}
\caption{\red{Latency vs. the number of coroutines.}}
\vspace{-4ex}
\label{fig:eval:coros}
\end{figure}

\red{The coroutines within compute node threads are designed to yield upon dispatching a request and resume operation upon receiving responses from two-sided RPCs. The default setup of \sys uses two coroutines per thread, but we extend our evaluation to explore the influence of one or more per thread to ascertain the optimal configuration for maximizing server CPU utilization.
Fig.~\ref{fig:eval:coros} studies the latency-throughput performance of \sys in YCSB-C workload with different numbers of coroutines in a compute node thread. 
In Fig.~\ref{fig:eval:cores:a}, we have only one worker thread in the memory node and vary the total of compute node threads as 8,20,72,144 and 216 distributed among three compute nodes, respectively. 
We can observe that a larger number of coroutines results in higher throughput when the number of compute node threads is less than 72, and the latency doubles or triples after the throughput reaches around 6 Mops, the maximum throughput one memory thread can support. 
This phenomenon is similar when the number of memory node threads is 2, as shown in Fig.~\ref{fig:eval:cores:b}, because the CPU resource on the memory node can handle 144 compute node threads, and the total throughput of a memory node can reach to 9.89 Mops. 
However, the extra coroutines will incur high latency of the data query after the number of memory node threads becomes a bottleneck for serving 216 threads.}

\vspace{-1.5ex}
\subsection{Influence of load factor in DMPH}
\label{sec:eval:lf}

\begin{figure}[!t]
    \renewcommand\thesubfigure{}
    \begin{minipage}[t]{0.232\textwidth}
        \subfigure[]{
            \label{fig:eval:lf}
            \includegraphics[width=\textwidth]{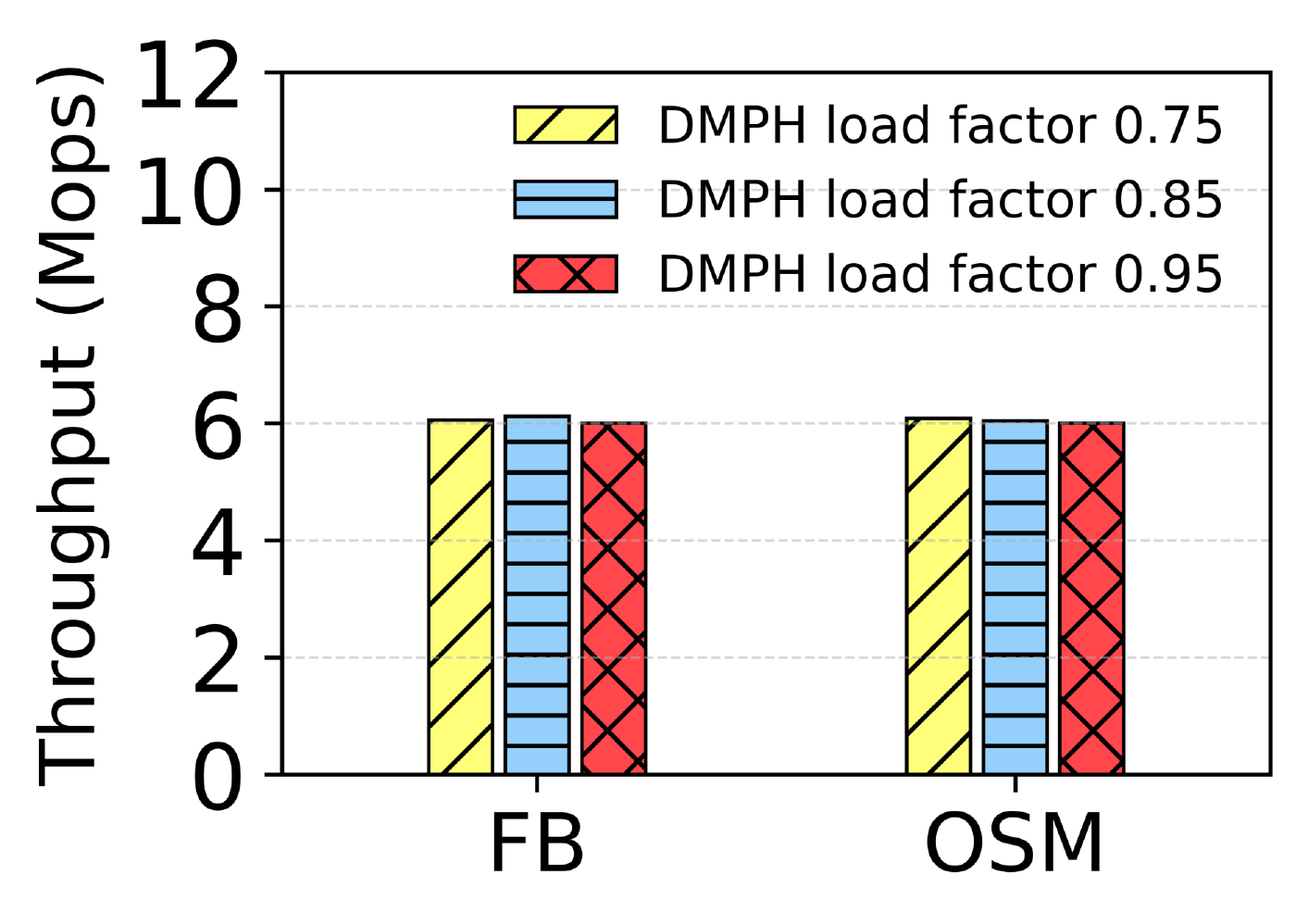}}
        \vspace{-7ex}
        \caption{Influence of different load factor set in DMPH.}
    \end{minipage}
    \hspace{-1.5ex}
    \begin{minipage}[t]{0.232\textwidth}
        \subfigure[]{
            \label{fig:eval:kvs}
            \includegraphics[width=\textwidth]{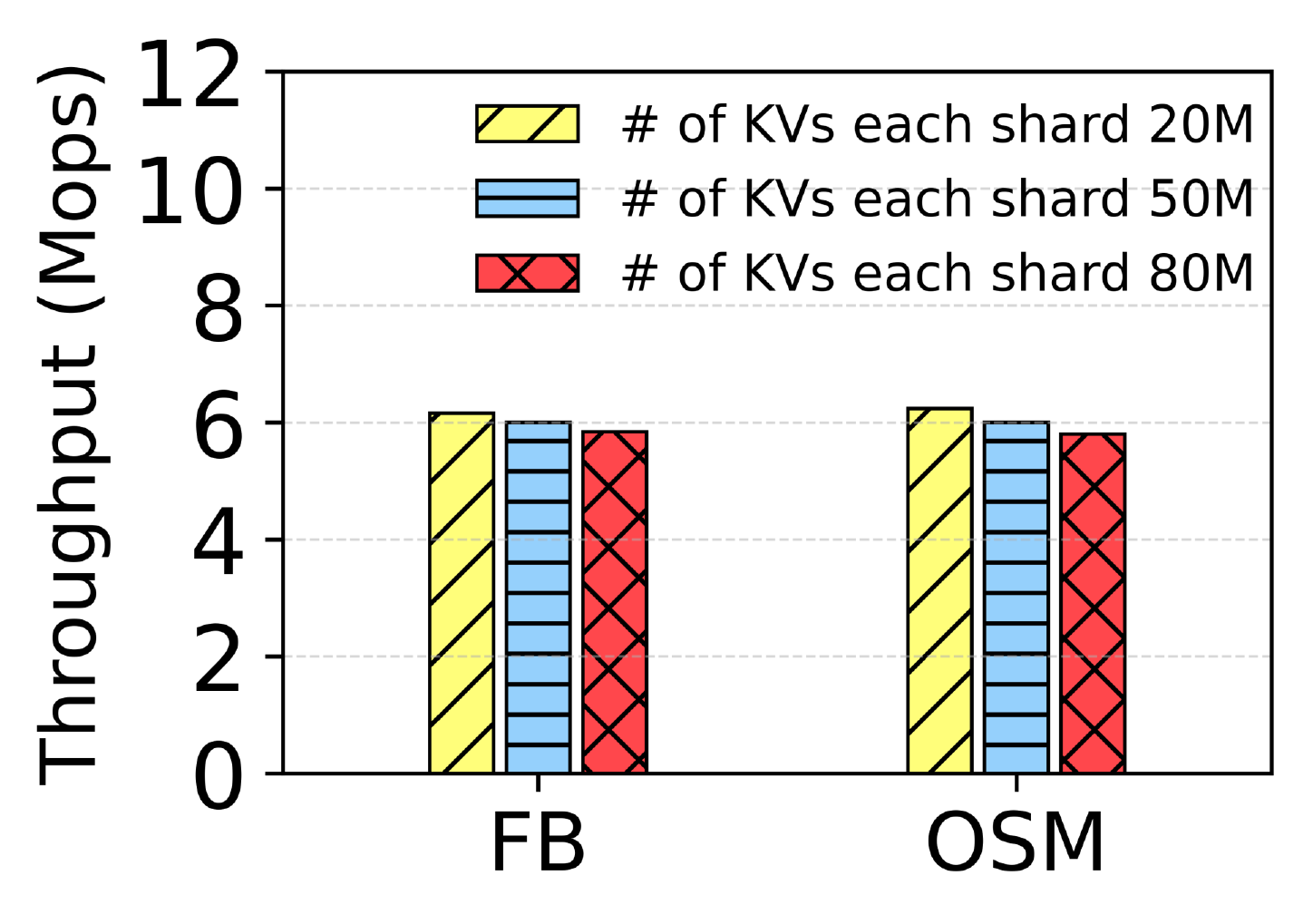}}
        \vspace{-7ex}
        \caption{Influence of the varied number of KV pairs.}
    \end{minipage}
    \vspace{-3ex}
\end{figure}

\red{The load factor in a hash table is the ratio of stored elements to the total number of available slots or buckets. Maintaining an optimal load factor balances memory usage and data operation throughput.
We evaluate the data \texttt{Get} throughput in \sys with varied load factors from 0.75 to 0.95.}

\red{As shown in Fig.~\ref{fig:eval:lf}, \sys can achieve around 6 Mops with 72 data query threads from compute nodes in a shard for the dataset FB.
Similarly, the influence of the varied load factors on the throughput is trivial based on the results of the dataset OSM.}

\vspace{-1.5ex}
\subsection{Influence of the number of KV pairs}
\label{sec:eval:kvs}
\red{Fig.~\ref{fig:eval:kvs} studies the impact of the number of KV pairs in each shard. We load 20M, 50M, and 80M KV pairs in \sys and evaluate the data \texttt{Get} throughput on two real-world datasets, respectively. 
\sys's read throughput decreases from 6.02 to 5.83 Mops as database size enlarges on the dataset FB. Similarly, we can observe the data read throughput decreases by 3.1\% on the dataset OSM.}

\vspace{-1.5ex}
\subsection{Memory usage in compute nodes}
\label{sec:eval:mem}

\begin{figure}[t]
\centering
\hspace{-.5ex}
    \includegraphics[width=0.45\textwidth]{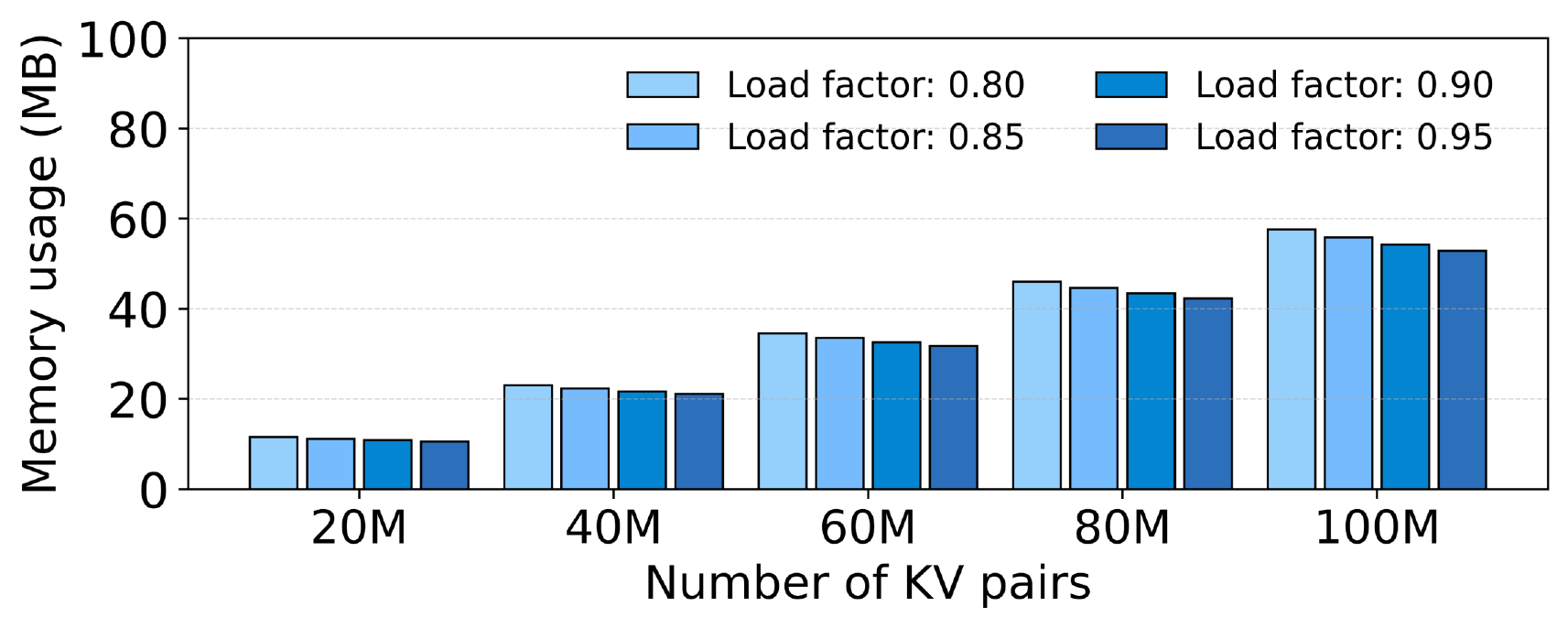}
    \vspace{-3ex}
    \caption{Memory usage on compute node with the varied number of KV pairs.}
    \label{fig:eval:mem}
    \vspace{-4ex}
\end{figure}

In a disaggregated memory system, compute nodes are regarded as the ones with rich computing resources but limited memory space. 
To make the memory node serve data requests with the least computation based on RDMA RPC primitives, we offload as much computation to the compute side with the help of DMPH.
In this section, we evaluate the memory cost of \sys on each compute node with the varied number of KV pairs in each shard. The memory usage on a compute node consists of the bucket locator and the seeds array.

As shown in Fig.~\ref{fig:eval:mem}, we vary the load factor used in the DMPH table from 0.80 to 0.95, and we use an 8-bit seed for keys in each bucket. The memory usage at each compute node for 20 million KV pairs per shard is around 12.5MB, and the cost is below 60MB for 100M KV pairs per shard.  
This is considered a small overhead because recent one-sided RDMA solutions cost hundreds of MBs or more on each compute node for index caching and other purposes~\cite{rolex,xtore}. 
For example, in XStore~\cite{xtore}, 100 million key-value pairs require over 600MB of memory at a compute node without including the cache.

\vspace{-1.5ex}
\subsection{Throughput during index resizing}
\label{sec:eval:resizing}

\begin{figure}[t]
\centering
    \includegraphics[width=0.45\textwidth]{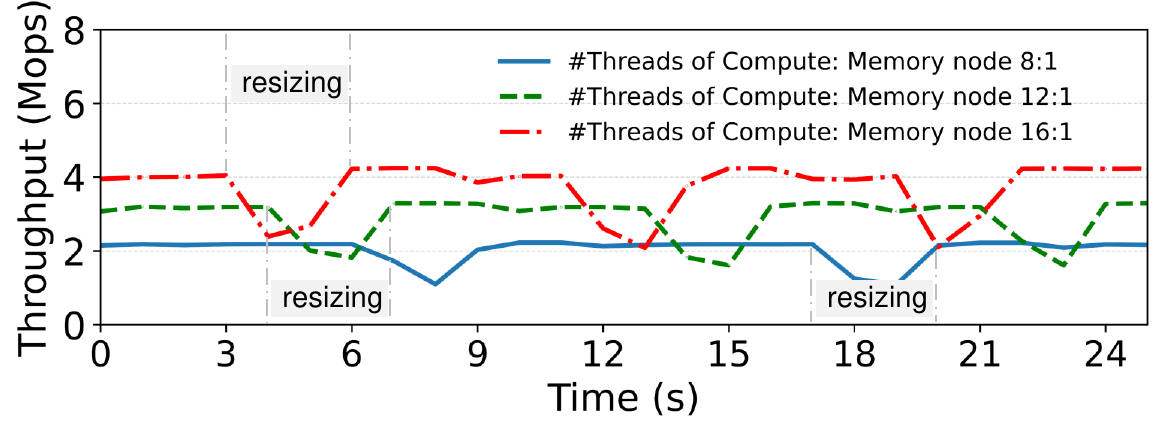}
    \vspace{-3ex}
    \caption{\red{Influence of extendible hashing resizing.}}
    \label{fig:eval:resizing}
    \vspace{-4ex}
\end{figure}

We evaluate the throughput changes during index reconstruction and resizing. 
\red{
In this set of experiments, we bulk-load 20M keys to the database with the initial DMPH table to warm up, and we set one compute node with 8, 12, and 16 threads connecting to the memory node running only one thread, respectively.} 
\red{This emulates a challenging scenario because the memory node has limited computing resources to handle both resizing and lookups. The workload running on compute nodes is YCSB D, which contains 5\% data insert and 95\% read.
As shown in Fig.~\ref{fig:eval:resizing}, it takes around 3 seconds to recalculate the bucket locator and the seed for each bucket. \sys still supports partial \texttt{Get} requests during resizing with a decreased throughput by approximately 52\% with only one thread in the memory node. 
The CPU contention causes a performance drop, 
and the performance goes back to normal after resizing.}

\vspace{-1.5ex}
\subsection{Summary of evaluation}
\label{sec:eval:sum}
\red{
\textbf{Data lookup throughput. }
\sys achieves 1.11-2.43$\times$ and 1.23-5.03$\times$ higher throughput than baselines with Mellanox CX-6 100Gb and CX-3 50Gb RNICs in data search workload, respectively.}

\red{\noindent\textbf{Memory usage. }
The memory usage at each compute node for 20 million KV pairs per shard is around 12.5MB per shard, around 5 bits per key, with a load factor of 0.85 in DMPH.}

\red{\noindent\textbf{Scalability of memory node threads.}
When the compute nodes with enough threads exhaust the compute capability on the memory node, \sys can achieve at least 18\% performance advantage over other RPC-based baselines on read workload.}

\red{\noindent\textbf{Load factors in \sys.}
The load factor value in DMPH causes a trivial impact on data lookup throughput with the same compute complexity. We recommend 0.8-0.9 to achieve the balance between memory usage and low frequent resizing, as the low load factor supports more incremental data insertion into the hash table.}

\vspace{-1.5ex}
\section{Related Work}
\label{sec:related_work}
\vspace{-.5ex}

\textbf{RDMA-based storage systems.} 
Existing RDMA-based storage can be classified into one-sided RDMA, RPC, or hybrid methods.
One-sided RDMA-based approaches~\cite{drtmr,farm,rolex,race,smart,cowbird} can bypass the memory node's CPU, managing data by RDMA\_READ, RDMA\_WRITE and other atomic verbs. 
Two-sided RDMA-based schemes~\cite{fasst,guidelines,reflex,stingray,smash} need only one round trip but suffer from the remote CPU bottleneck, posing challenges in saturating RNIC bandwidth due to the computation burden for the callback data service. The index data structures of existing two-sided RDMA, such as hash table~\cite{pilaf,mica}, learned index~\cite{finedex, rolex} and Blink Tree~\cite{kraska}, put the memory node's CPU in charge of nontrivial computation tasks. 
The hybrid methods~\cite{cell,drtmh,herd,hstore} combine two of the above approaches to boost the throughput. 


In addition to examining design primitives and communication protocols within RDMA-based systems.
Cowbird~\cite{cowbird} frees the CPU burden in compute nodes by offloading RDMA posting tasks on in-network computation devices (e.g. programmable switch~\cite{netcache}), so that the compute node can focus on computation duties. 
SmartNIC~\cite{smartnic1,smartnic2,prism,strom} can also be put in the network interface and works as an extra compute core on the critical data path, and it enables compute nodes to access data without network or RPC overhead. 
Note that the computation resource required in memory nodes of \sys can also be offloaded to SmartNIC or SmartSSD, whose SOCs are closer to data. 

\textbf{Minimal perfect hashing for networked systems.} 
Perfect hashing offers a rapid method for data indexing, effectively preventing hash collisions. Moreover, DMPH enhances memory efficiency by eliminating the need to store keys and mapping $N$ elements into $(1+\epsilon)N$ space within the table.
Besides the Ludo hashing shown in \S ~\ref{sec:background}, Setsep~\cite{scalebricks} leverages a novel two-level hashing scheme that distributes billions of keys across cluster servers with a memory cost of $0.5+1.5l$ bits/key. 
BuRR~\cite{burr} is another MPH scheme that involves manipulating a matrix for each key, and the multiplication values of keys determine various ranks within the bucket.

\vspace{-1.5ex}
\section{Conclusion}
\label{sec:conclusion}
\vspace{-.5ex}
This paper introduces \sys, an RDMA RPC-based index for key-value stores on disaggregated memory, designed to achieve high throughput with lower CPU utilization. 
The key innovation of \sys is the division of the data index into two distinct components: a compute-intensive component cached on compute nodes and a memory-intensive component residing on memory nodes. The performance improvements stem from the memory node's ability to access underlying data with minimal computational overhead with perfect hashing.
We also design protocols for \sys that support data operations and index resizing using extendible hashing, ensuring both the correctness of operations and system consistency during updates.
We conduct extensive experiments to evaluate the performance of \sys. The results show that \sys achieves higher throughput and requires smaller memory space on compute nodes, compared to the state-of-the-art baselines under most types of workload, especially for $\mathtt{Get}$-heavy workload. 

\vspace{-2ex}
\begin{acks}
\vspace{-1ex}
We thank our three anonymous reviewers for their insightful suggestions and comments.
This research was supported by the IAB members of the Center for Research in Systems and Storage (CRSS), and the National Science Foundation (NSF) under grants CNS-1841545, CCF-1942754, CNS-2322919, CNS-2420632, CNS-2426031, and CNS-2426940. The views expressed are those of the authors and do not necessarily reflect those of the funding agencies.
\end{acks}

\balance
\bibliographystyle{ACM-Reference-Format}
\bibliography{sample-base}

\end{document}